\documentclass[aps, reprint,
        10pt, notitlepage,
        floats, floatfix,
        amsmath, amssymb, amsfonts,
        superscriptaddress,
        showpacs, showkeys,
        nofootinbib,prd
]{revtex4-2}
\usepackage{aasmacros}
\usepackage{mathtools}
\usepackage{graphicx} 
\usepackage[caption=false]{subfig} 
\usepackage{xspace} 
\usepackage[utf8]{inputenc} 
\usepackage{verbatim}
\usepackage[usenames,dvipsnames]{xcolor}
\usepackage[colorlinks = true,
            linkcolor = blue,
            urlcolor  = blue,
            citecolor = blue,
            anchorcolor = blue]{hyperref}
\usepackage[capitalise]{cleveref}
\crefname{section}{Sec.}{Secs.}
\Crefname{section}{Sec.}{Secs.}
\crefname{algocf}{Alg.}{Algs.}
\Crefname{algocf}{Alg.}{Algs.}
\usepackage{orcidlink}
\usepackage{colortbl,bm,booktabs,multirow}
\usepackage[normalem]{ulem} 

\newcommand{\ve}[1]{\boldsymbol{#1}}

\newcommand{\code}[1]{\texttt{#1}\xspace}
\newcommand{\gpry}{\code{GPry}}
\newcommand{\nessai}{\code{nessai}}
\newcommand{\cobaya}{\code{Cobaya}}
\newcommand{\cosmomc}{\code{CosmoMC}}
\newcommand{\balrog}{\code{Balrog}}

\newcommand{\msun}{\mathrm{M}_\odot}

\newcommand{\djs}{D_{\scriptscriptstyle \mathrm{JS}}}
\newcommand{\dkl}{D_{\scriptscriptstyle \mathrm{KL}}}
\newcommand{\dwd}{DWD\xspace}
\newcommand{\dwds}{DWDs\xspace}
\newcommand{\sbhb}{stBHB\xspace}
\newcommand{\sbhbs}{stBHBs\xspace}
\newcommand{\smbhb}{SMBHB\xspace}
\newcommand{\smbhbs}{SMBHBs\xspace}

\begin{document}

\title{Accelerating LISA inference with Gaussian processes}

\newcommand{\bham}{\affiliation{Institute for Gravitational Wave Astronomy \& School of Physics and Astronomy, University of Birmingham, Birmingham, B15 2TT, UK}}
\newcommand{\insubria}{\affiliation{Como Lake Center for Astrophysics, Department of Science and High Technology, University of Insubria, via Valleggio 11, I-22100, Como, Italy}}
\newcommand{\Bic}{\affiliation{Dipartimento di Fisica “G. Occhialini”, Università degli Studi di Milano-Bicocca, Piazza della Scienza 3, 20126 Milano, Italy}}
\newcommand{\Infn}{\affiliation{INFN, Sezione di Milano-Bicocca, Piazza della Scienza 3, 20126 Milano, Italy}}
\newcommand{\stavanger}{\affiliation{Department of Mathematics and Physics, University of Stavanger, NO-4036 Stavanger, Norway}}
\newcommand{\unipadova}{\affiliation{Dipartimento di Fisica e Astronomia “G. Galilei”, Università degli Studi di Padova, via Marzolo 8, I--35131 Padova, Italy}}
\newcommand{\infnpadova}{\affiliation{INFN, Sezione di Padova, via Marzolo 8, I--35131 Padova, Italy}}
\newcommand{\iemcsic}{\affiliation{Instituto de Estructura de la Materia, CSIC, Serrano 121, 28006 Madrid, Spain}}

\author{Jonas El Gammal~\orcidlink{0009-0005-4833-4266}}
\email{jonas.e.elgammal@uis.no}\stavanger\insubria
\author{Riccardo Buscicchio~\orcidlink{0000-0002-7387-6754}}
\Bic
\Infn
\bham
\author{Germano Nardini~\orcidlink{0000-0002-3523-0477}}
\stavanger
\author{Jesús Torrado~\orcidlink{0000-0002-5194-1877}}
\unipadova\infnpadova\iemcsic

\date{\today}

\begin{abstract}
Source inference for deterministic gravitational waves is a computationally demanding task in LISA.
In a novel approach, we investigate the capability of Gaussian Processes to learn the posterior surface in order to reconstruct individual signal posteriors. We use \texttt{GPry}, which automates this reconstruction through active learning, using a very small number of likelihood evaluations, without the need for pretraining.
We benchmark \texttt{GPry} against the cutting-edge nested sampler \texttt{nessai}, by injecting individually three signals on LISA noisy data simulated with \texttt{Balrog}: a white dwarf binary (DWD), a stellar-mass black hole binary (stBHB), and a super-massive black hole binary (SMBHB).
We find that \texttt{GPry} needs $\mathcal O(10^{-2})$ fewer likelihood evaluations to achieve an inference accuracy comparable to \texttt{nessai}, with Jensen-Shannon divergence $D_{\scriptscriptstyle \mathrm{JS}} \lesssim 0.01$ for the DWD, and $D_{\scriptscriptstyle \mathrm{JS}}  \lesssim 0.05$ for the SMBHB. Lower accuracy is found for the less Gaussian posterior of the stBHB: $D_{\scriptscriptstyle \mathrm{JS}} \lesssim 0.2$.
Despite the overhead costs of \texttt{GPry}, we obtain a speed-up of $\mathcal O(10^2)$ for the slowest cases of stBHB and SMBHB.
In conclusion, active-learning Gaussian process frameworks show great potential for rapid LISA parameter inference, especially for costly likelihoods, enabling suppression of computational costs without the trade-off of approximations in the calculations.
\end{abstract}

\maketitle

\section{\label{sec:intro}Introduction}

In the last decade, the direct detection of gravitational waves (GWs) has transformed from a remarkable, singular accomplishment into a routine procedure.
Currently, the LIGO-Virgo-KAGRA collaboration has observed approximately a hundred systems emitting GWs in the $10$ -- $1000\,\mathrm{Hz}$ frequency range~\cite{KAGRA:2021vkt}.
Moreover, pulsar timing array experiments are possibly on the verge of gathering enough statistics to announce the first direct detection of GWs in the nHz range~\cite{NANOGrav:2023hvm, EPTA:2023fyk, Reardon:2023gzh, Xu:2023wog}.
Gravitational waves in the mHz frequency range remain unobserved.
LISA, with construction commissioned now and launch scheduled in a decade, is set to delve into this uncharted territory~\cite{LISA:2017pwj}.

LISA poses data analysis challenges that are radically different from those of the other GW experiments, as it is a signal-dominated one, expected to observe a multitude of Galactic binaries, supermassive BHBs, EMRIs, and stellar-mass BHBs constantly populating the datastreams~\cite{LISA:2017pwj}.
A primordial stochastic gravitational wave background (SGWB) as loud as the astrophysical sources might also be present~\cite{LISACosmologyWorkingGroup:2022jok}.
To further complicate things, the zoology of LISA signals does not admit a common detection and reconstruction strategy: within the experiment's lifetime, some sources are monochromatic, others slowly drift, and others move fast outside the LISA frequency sensitivity.
Analyzing the data in time chunks makes the identification of the long-duration sources more difficult, while keeping the whole datastream a priori makes the likelihood evaluations too heavy.
On the other hand, splitting the data into frequency intervals is suitable for monochromatic sources \cite{Strub:2022upl}, less so for those that are largely chirping.
Despite their difficulty, these challenges must be solved to achieve the groundbreaking science promised by LISA~\cite{LISA:2022yao, LISAConsortiumWaveformWorkingGroup:2023arg, LISACosmologyWorkingGroup:2022jok}.

Concerning the likelihood evaluation cost, several improvements are conceivable (see e.g., \cite{Cuoco:2020ogp} for a review):
speeding up waveform evaluation (e.g., through hardware acceleration or approximation schemes) \cite{Canizares:2014fya,Smith:2016qas,Field:2013cfa}, by-passing the likelihood evaluation (e.g., using simulation-based inference methods (SBI) \cite{Dax:2021tsq,Bhardwaj:2023xph,Andres-Carcasona:2023rnk,Chatterjee:2024pbj,Dax:2024mcn,Raymond:2024xzj,Vilchez:2024qnw,sun2025acceleratingbayesiansamplingmassive}), or building surrogates of the likelihood function itself \cite{Gabbard:2019rde,Dax:2022pxd}. In this paper, we focus on the latter, while agnostically retaining the waveform content and the signal Bayesian model intact.
Thus, no approximation is made to either the Fourier transforms of the modeled signals or the likelihood computation.

Instead, we adopt the machine learning framework implemented in \gpry \cite{gpry_1,gpry_2}.
Within it, we interpolate the posterior with a Gaussian process \cite{gpml},
trained on a small number of evaluations that are sequentially proposed in an optimal way to minimize their number \cite{ALoMEuBQ,Henning_inference}.
As we will see, this approach can produce accurate inference with $\mathcal{O}(10^{-2})$ fewer likelihood evaluations than traditional Monte Carlo approaches.
This translates into a speed-up of inference by a factor of 100 in the regime in which the overhead of \gpry is subdominant, i.e., when the likelihood evaluation time is over a few seconds, and the dimensionality of the problem is $\mathcal{O}(10)$.
The output is a surrogate model for the posterior that can be sampled with a Monte Carlo algorithm at virtually zero cost.

Within the general context of machine-learning accelerated inference of GW sources, the likelihood-based, \textit{active-learning} approach taken by \gpry differs from the likelihood-free, \textit{amortized} approaches such as SBI in a number of ways:
a) amortized approaches are much faster at the point of inference, in exchange for some costly pre-training, whereas the more expensive active-learning frameworks can be run with no upfront costs for variations of data or waveform modelling;
b) likelihood-based approaches do not necessitate the simulated data to contain stochastic noise, and possess a direct way to evaluate goodness-of-fit.
Both approaches are complementary and can thus coexist in the LISA parameter inference pipeline. To estimate the benefits for LISA of a machine learning framework similar to the one just described, we use \gpry to perform parameter inference on some benchmark LISA signals simulated through \balrog, and compare the speed and accuracy of the results to those obtained with the state-of-the-art nested sampler \nessai.

The paper is organized as follows:
in \cref{sec:source_types} we describe the target sources of our study: a supermassive black-hole binary (\smbhb), a stellar mass black hole binary (\sbhb) and a Galactic double white dwarf (\dwd) system;
in \cref{sec:data} we compare the waveform modeling available in literature;
in \cref{sec:likelihood} we briefly describe the inference scheme, for individual source parameter estimation in the three source scenarios previously mentioned;
in \cref{sec:searches} we present previous approaches for exact or approximate inference, and how they can be used as a starting point for our pipeline;
in \cref{sec:gp} we briefly introduce how our algorithm models a posterior using a Gaussian process interpolator;
in \cref{sec:inf} we detail how we perform, evaluate and compare our inference runs;
in \cref{sec:results} we present our results for the three source types above, and compare \gpry and \nessai on performance and accuracy;
in \cref{sec:conclusions} we draw conclusions and outline possible future developments.

\section{\label{sec:astro}Sources}
\subsection{\label{sec:source_types}Source types}
We explore three different source classes, roughly categorized by their signal spectral content.
Double white dwarfs (\dwds) are observed by LISA during their early inspiral, emitting quasi-monochromatic GWs largely detectable within the Galactic neighborhood~\cite{2023MNRAS.519.2552G,2021MNRAS.502L..55K,2024arXiv240207571C}.
Each signal persists in the LISA datastream for the entire mission, Doppler modulated by the satellite-constellation orbital motion within a very narrow frequency band ($\Delta f/f \leq 10^{-4}$).
For \dwds emitting above approximately $2\,{\rm mHz}$, the GW-driven orbital tightening reaches a frequency evolution $\dot{f} \gtrsim 10^{-15}\,{\rm Hz}^2$ which LISA can measure over the nominal mission duration $T_{\rm LISA}=4\, {\rm yr}$.
As many as $10^7$ \dwd sources are expected to emit in the LISA band, with up to $1\%$ individually detectable.
They are unambiguously the most numerous deterministic sources expected for LISA, and their collective brightness makes its datastream strongly signal dominated below a few mHz.
The brightest of these sources are identifiable after a few months~\cite{2023MNRAS.522.5358F}, once a sufficient phase coherence emerges from the noisy datastream.
In most of the available literature, a phenomenological parametrization of their signal is preferred over a physically-motivated one.
Waveform models accurately taking into account the LISA response~\cite{1998PhRvD..57.7089C,GPGPU} are typically fast, often leveraging frequency domain representation and heterodyning.
It is uncertain whether such advantages will be retained in more realistic data analysis setups (see, e.g., \cite{2025arXiv250217426B}, for recent developments on a frequency domain treatment of gaps).

\sbhbs are the second class of sources we consider. They are expected to populate the whole LISA spectrum, the largest majority slowly drifting in frequency within the LISA mission duration. Only a handful of them will exit the band on its upper end in less than a year and eventually merge in the ground-based detector frequency band ($10\, {\rm Hz}$ to $1\, {\rm kHz}$) \cite{Buscicchio:2024asl,2022arXiv220403423K}.
Their waveforms are comparatively more complex than the \dwd ones, with a parameter space equipped to describe eccentric, precessing, unequal-mass binaries~\cite{2021arXiv210610291K,2025arXiv250203929M}.
The in-band persistence of \dwd and \sbhb signals allows for a coherent integration of the data over millions of cycles during the nominal mission duration, making their detection heavily phase dominated.

Finally, we consider \smbhbs as the third category of GW sources: they are the most massive binaries expected to emit GWs in the LISA band.
In the lifetime of the mission, \smbhbs will be detected as transient signals reaching signal-to-noise ratios (SNRs) as large as $\sim 10^3$, therefore being the loudest individual sources among the LISA ones.
\smbhbs rapid evolution towards merger in band makes them prototypical to excite GW higher multipole modes, spin-precession~\cite{2023PhRvD.108l4045P}. However, orbit circularization~\cite{PhysRev.131.435} prevents from measuring large orbital eccentricities.
State-of-the-art waveform models are phenomenological ones, calibrated against numerical relativity simulations with mass-ratios up to 1:18~\cite{2020PhRvD.102f4002G}.
Their computational efficiency is granted by decades of waveform developments for ground-based detectors, though the signal brightness questions the level of accuracy required to achieve unbiased parameter estimation~\cite{2023arXiv231101300L}.
The broadband nature of \smbhbs waveforms makes them the most expensive to compute in frequency domain (only second to extreme mass ratio inspirals), with up to $10^5$ datapoints required for the lightest, most distant sources merging at about $10\, {\rm mHz}$.
Even though time-domain truncation may reduce the number of frequencies to evaluate the waveform at, advanced global inference schemes (e.g. Gibbs-like sampling or SBI techniques) may require the usage of conditional data with full-resolution frequency series.

\subsection{\label{sec:data}Signal model}
We model LISA data $d$ as the linear superposition of noise $n$ and signal $s$.
Observations are collected through time-delay-interferometric variables, synthetic time series constructed from suitable delayed combinations of single-link inter-spacecraft laser phase measurements~\cite{2005LRR.....8....4T}.
For simplicity, we assume the three LISA satellites orbiting in an equilateral triangular configuration with constant armlength of $2.5\times 10^9\, {\rm m}$.
Under such an approximation, the three interferometric variables, often referred to in literature as $X,Y,Z$, are linearly combined into the $A,E,T$ variables, such that the respective noises are uncorrelated.

We model the GW strain emitted from a distant DWD as a quasi-monochromatic signal.
Its two polarizations are described by
\begin{align}
    h_+(t; \boldsymbol{\theta}) &= A (1+\cos^2 \iota) \cos (2 \pi f_\mathrm{GW}(t)/{\rm Hz} - \phi)\,,\\
    h_\times (t; \boldsymbol{\theta}) &= -2A (\cos \iota) \sin (2 \pi f_\mathrm{ GW}(t)/{\rm Hz} - \phi)\,,
\end{align}
where ${A}$ denotes the GW amplitude, $\iota$ represents the source inclination with respect to the line-of-sight, $f_{\rm GW}(t)$ is the instantaneous GW frequency measured in the solar system barycenter frame, and $\phi$ is the binary orbital phase at the time $t_0$ at which LISA observations start.
The amplitude $A$ can be expressed as
\begin{equation}
    {A} = \frac{2(G \mathcal{M}_c)^{5/3}}{c^4 d_L} (\pi f)^{2/3}, \label{eq:GWamplitude}
\end{equation}
while, to leading order, $f_{\rm GW}(t)$ reads
\begin{equation}
    f_{\rm GW}(t)=f + \dot{f} (t-t_0)\,,
\end{equation}
with $f$ and $\dot{f}$ being the orbital frequency and its (solar system barycenter frame) time derivative at time $t_0$, respectively.
In \cref{eq:GWamplitude}, $d_L$ denotes the source luminosity distance whose redshift is $z$, and
${\cal M}_c$ denotes the chirp mass
\begin{equation}
  \label{eq:chirpmass}
  \mathcal{M}_c = \frac{(m_1 m_2)^{3/5}}{(m_1 + m_2)^{1/5}} \,,
\end{equation}
for a binary system of two component masses $m_1$ and $m_2$. \cref{eq:chirpmass} is frame-invariant, however we consider only solar system barycenter frame quantities hereafter.

The LISA detector response introduces an additional dependence upon the source position in the sky.
We parametrize it by the source Ecliptic latitude $b$ and longitude $\lambda$, and an overall polarization angle $\psi$.
For inference purposes, we also reparameterize $\phi, \psi$ with two circular initial phases $\phi_L = \phi + \psi$ and $\phi_R = \phi - \psi$, respectively.

We assume the DWD source orbital evolution to be GW driven when injecting it into LISA data.
Therefore, the injected $f$ and $\dot{f}$ must satisfy the constraint
\begin{equation}
    \dot{f} = \frac{96}{5}\frac{(G {\cal M}_c)^{5/3}}{\pi c^5} (\pi f)^{11/3}\,.
\end{equation}
However, we infer $f$ and $\dot{f}$ as free independent parameters.
Due to the signal being narrowband and at a much lower frequency than the LISA data sampling rate ($f_s= 0.2\, {\rm Hz}$), we speed up likelihood evaluations through heterodyning, filtering, and downsampling, resulting in a few hundred of datapoints per waveform evaluation.

Concerning \sbhbs, we model their GW signal by following~\cite{2021PhRvD.104d4065B} where the waveform is computed through an adiabatic inspiral post-Newtonian expansion.
As \sbhbs drift much faster than \dwds in the LISA frequency band, their waveform exhibits (mild) sensitivity to the component masses $m_1, m_2$ and dimensionless spins $\chi_1, \chi_2$.
For simplicity, we consider aligned-spin systems in circular orbits only, leaving the investigation of eccentric, precessing ones for future work.
We reduce inference correlations with a convenient physical parameterization through the binary chirp mass ${\cal M}_c$, reduced mass ratio $\delta\mu = (m_1 -m_2)/(m_1 + m_2)$, and component dimensionless spin magnitudes $\chi_{1,2}$; its initial orbital frequency $f_0$, and left- and right-handed phases $\phi_L,\phi_R$. The extrinsic parameters are decomposed as follows:
the source position and inclination are parameterized by the square root of two circular amplitudes $A_{L,R} = (1 \pm \cos\iota)/\sqrt{2 d_L}$, the sin-ecliptic latitude $\sin \beta$, and longitude $\lambda$.
The values for the time-delay interferometry (TDI) quantities \cite{Babak:2021mhe} are constructed through a rigid adiabatic approximation~\cite{2004PhRvD..69h2003R}.
In previous work~\cite{2021PhRvD.104d4065B}, waveforms were evaluated only at a few hundreds of points, employing Clenshaw-Curtis quadrature to approximate the likelihood in \cref{eq:loglikelihood}. This was made possible by analyses of noiseless data, whose smoothness allows for such an integration scheme.
In turn, in this work we focus on noisy data, and hence we use the full GW frequency content, resulting in around $10^4$ data points per waveform.

Finally, \smbhbs signals are described through phenomenological, numerical-relativity calibrated waveforms, as implemented in \texttt{IMRPhenomXHM}~\cite{2021PhRvD.103j4056P}.
This waveform family smoothly captures the inspiral-merger-ringdown structure of a binary merger signal in frequency domain, accounting for higher-modes emission.
Despite being extremely fast, thanks to decades-long optimization for current and future ground-based detectors~\cite{2020PhRvR...2b3151P}, the LISA frequency resolution makes the waveform array typically long:
in this study we consider a system emitting up to $3.5\,\textrm{mHz}$, reaching its merger $\tau_m=4.138 \times 10^6 {\rm s}$ after the start of the mission.
We do not consider any time-domain truncation scheme, and model the signal at the highest frequency resolution available.
The signal is parameterized by the binary chirp mass, its reduced mass ratio, the component dimensionless spin magnitudes (assumed aligned with respect to the angular momentum), the time-to-merger $\tau_m$,  the luminosity distance $d_L$, the sine-ecliptic latitude $\sin \beta$ and ecliptic longitude $\lambda$, the cosine inclination $\cos \iota$, the initial orbital phase $\phi_0^{\rm orb}$, and the polarization angle $\psi$.
All quantities are defined in the solar system barycenter frame, and non-conserved ones are defined at a reference frequency $f_{\rm ref} = 10^{-4}\, {\rm Hz}$.

Throughout this work, we assume perfectly known, Gaussian, instrumental noise~\cite{LISA2018}, superimposed on likewise perfectly known, Gaussian, confusion noise, whose level is modeled as in~\cite{2021PhRvD.104d3019K} as a function of $T_{\rm LISA}$.
In the larger context of global fit pipelines, this is equivalent to performing inference on the three chosen sources after all resolvable ones have been identified and perfectly subtracted from the data.
We further simplify the two noise models assuming both zero mean and perfectly stationary, thus reducing their entire description to simple power spectral densities~\cite{LISA2018}.

\subsection{\label{sec:likelihood}Likelihood}
Given the assumptions detailed in \cref{sec:data}, the likelihood of observed data $d_k =A,E,T$ in frequency domain reads
\begin{equation}
    \log {\cal L}(d | {\boldsymbol{\theta}})\! = \!-\!\sum_k \frac{\left\langle d_k\! -\! s_k({\boldsymbol{\theta}}) \mid d_k\! -\! s_k({\boldsymbol{\theta}})\right\rangle_k}{2}\!+ {\rm const.}\label{eq:loglikelihood}
\end{equation}
where $d_k$ denotes the superposition of noises realizations and each injected signal as described in Table~\ref{tab:WD_params}, \ref{tab:SOBBH_params}, and \ref{tab:SMBBH_params}, respectively.
Thanks to the stationarity of noise in each datastream and uncorrelatedness across them, the inner product is simply given by
\begin{equation}
    \left\langle x \mid y \right\rangle_k = 4 \mathrm{ Re}\int_0^{+\infty} {\rm d}f \frac{\tilde{x}(f)\tilde{y}^\dagger(f)}{S_{n,k}(f)}\,.
\end{equation}
Finally, $s_k(f;\boldsymbol{\theta})$ denotes a proposed GW signal with parameters $\boldsymbol{\theta}$, as observed in the $k$-th datastream, and $S_{n,k}$ the noise power spectral density in the same datastream.
We characterize the overall source brightness with the $\mathrm{SNR}$, defined as
\begin{equation}
    \mathrm{SNR}^2=  \sum_{k=A,E,T} \left\langle s_k(f;\boldsymbol{\theta}) \mid s_k(f;\boldsymbol{\theta})\right\rangle_k  \,.\label{eq:SNR}
\end{equation}
In this study, we present two approaches to obtain posterior samples for each inference, according to
\begin{equation}
    \label{eq:posterior}
    p(\boldsymbol{\theta} | d) \propto {\cal L}(d | \boldsymbol{\theta}) \pi(\boldsymbol{\theta})\,,
\end{equation}
where $\pi(\boldsymbol{\theta})$ denotes the prior assumption $\boldsymbol{\theta}$. In \cref{sec:searches} we detail the construction of priors for each source category, which we assume to be uniform over the prescribed ranges.

\section{\label{sec:inference}Inference}
\subsection{\label{sec:searches}
Setting priors}
Pre-constraining the parameter space of the inference problem down to a region around the location of the bulk probability mass, henceforth referred to as ``mode'', makes surrogate-posterior approaches such as \gpry significantly faster and more robust. This can usually be achieved with methods that avoid the evaluation of the expensive posterior, via e.g.\ approximations in the likelihood, template matching with an approximate waveform or machine-learning forward modeling~\cite{Field:2013cfa,Canizares:2014fya,2025arXiv250209266S}. These methods can produce rough estimates of the location and span of the posterior mode at a very low computational cost.

The \dwds live in a narrow frequency band and can be initially constrained using frequentist triggers with a sliding-window method that scans the frequency domain. Additionally, by using an optimizer, it is possible to obtain a maximum likelihood estimate (MLE), and an estimate of the Fisher information matrix. In combination, these methods allow the setting of priors that sufficiently encapsulate the mode of the posterior distribution, as done in \cite{Strub:2022upl}.
Since the \dwd are mostly a test case for our study, we set conservative priors by hand, encapsulating $\sim 10\sigma$ for each unbounded parameter. 

or \sbhbs our approach to pre-constraining the parameter space is the one introduced in~\cite{Bandopadhyay_2023} and successfully applied to LISA data in~\cite{Bandopadhyay_2024}.
This method employs a semi-coherent search combined with Particle Swarm Optimization (PSO) to efficiently scan the large parameter space involved. The semi-coherent approach divides the data into frequency-domain segments, analyzing each individually, and then combining the results.
This technique balances sensitivity and computational efficiency by widening the posterior distribution over the parameter space, thus helping to locate the posterior bulk.

The path traced by the particles in the PSO can then be used to find regions in the parameter space with high posterior density values.
For our \sbhb analysis, we use a subset of 5000 samples from the PSO paths, obtained from 256 data segments, and evaluate the posterior in \cref{eq:posterior} at these locations.
We then restrict the prior to the smallest hyper-rectangle containing PSO posterior samples within a $10\sigma$ confidence region from the peak, assuming a multivariate Gaussian distribution as the posterior distribution (for a detailed discussion, see App.~A of \cite{gpry_1}). In addition,  we use a small set of these samples close to the top of the mode as an initial training set for \gpry.
Together, the shrunken prior and the initial training set eliminate the need to explore the parameter space and let \gpry focus on mapping the mode, thus combining the strengths of both approaches: a fast initial exploration of the parameter space by the PSO followed by \gpry which maps the mode with very few evaluations of the relatively slow-to-evaluate posterior distribution. The PSO search takes $\mathcal{O}(10\,\mathrm{min})$, adding only very little overhead to our pipeline.
We perform the initial PSO on noiseless data to introduce an additional bias beyond the one arising from their segmentation.

For the less explored case of \smbhbs, we assume that a similar PSO approach, a neural-network or a frequentist one can be used to approximate the mean and covariance of the posterior mode (see, e.g.,~\cite{2025arXiv250209266S,2025PhRvD.111d3045C}).

Due to the simpler structure of the posterior, we can set a larger uniform prior covering $>10\sigma$ in each dimension. As a proxy for a search, we generate Monte Carlo samples of the noiseless posterior distribution and use its mean and covariance to draw a set of 35 samples from a multivariate Gaussian distribution. \gpry is initialized with these samples which are close to the top of the mode but biased. If multimodalities are present, as is expected in the sky location for low latency searches~\cite{2023PhRvD.107l3026P,2023MNRAS.521.2577P,2022PhRvD.106j3017M}, \gpry would be initialized with points from all modes.

\subsection{\label{sec:gp}Gaussian process posterior interpolation}
\begin{figure*}[t!]
\centering
\includegraphics[width=\textwidth, keepaspectratio]{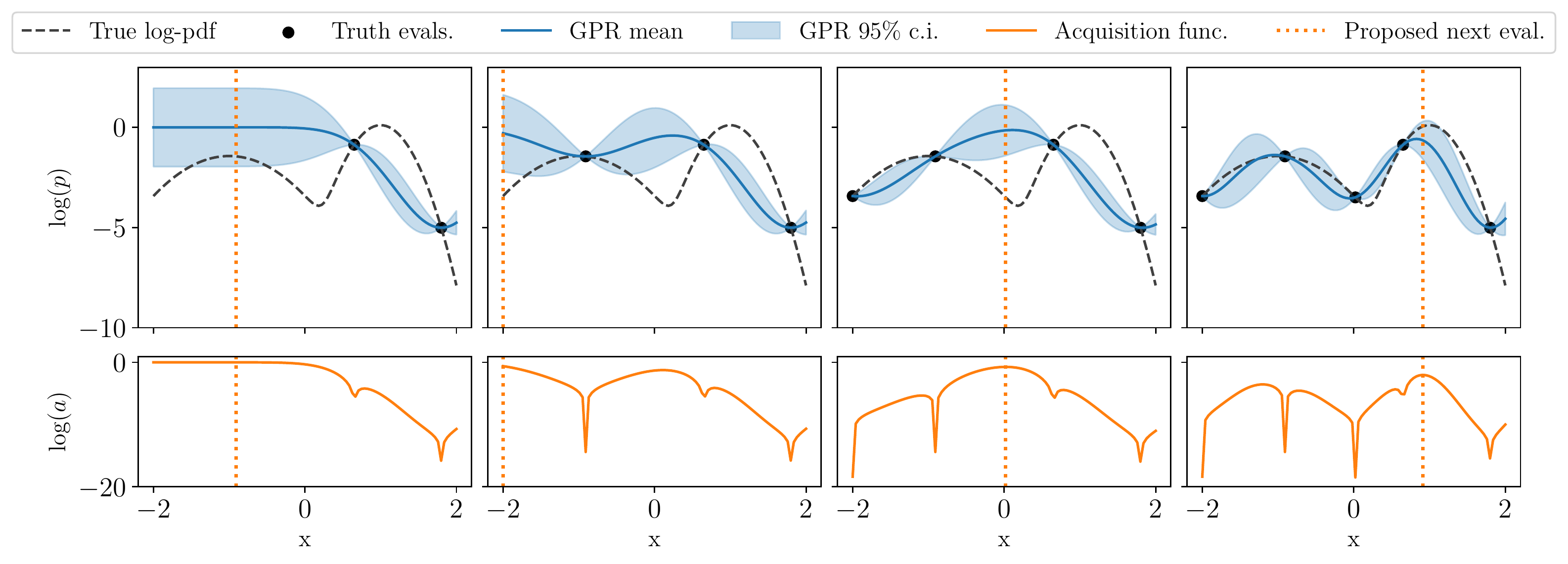}
\caption{
Simplified illustration of the \gpry algorithm on a 1-dimensional Gaussian mixture test function.
Each column, corresponding to consecutive iterations, shows on the top the true target log-pdf (dashed), the current set of evaluations (black points), and the current GPR model mean from \cref{eq:gp_mean} (blue, solid) and $95\%$ confidence interval defined by \cref{eq:gp_cov} (blue, shaded); the bottom panel shows the current acquisition function values from \cref{eq:acquisition_function}, whose maximum (dotted orange) will be proposed for evaluation for the next iteration.
Not illustrated are more complex aspects of the algorithm, such as the batch proposal of points \cite{gpry_1}, and the procedure to obtain approximate maxima of the acquisition function \cite{gpry_2}. 
}
\label{fig:likelihood-approximation_alt}
\end{figure*}

The inference algorithm employed in this study uses a Gaussian process regressor (GPR) to create an approximate model of the posterior density function, using a small set of evaluations performed at optimal locations. This approximation is then used as a \textit{surrogate} model from which we can draw, at very low computational cost, Monte Carlo (MC) samples that very closely resemble samples from the true posterior. Contrary to amortized machine learning-based approaches such as SBI, our surrogate model is built sequentially at runtime (an approach known as \textit{active learning}), and does not rely on previous training. \gpry's approach is more closely related to variational inference (see, e.g.,~\cite{Vallisneri:2024xfk}), with the difference that \gpry does not need derivatives of the posterior. As we will see, the necessary number of evaluations of the GW signal likelihood is at least $\mathcal{O}(10^{-2})$ smaller than those needed by \nessai (which is already more efficient than traditional Nested Sampling implementations).

We use \gpry \cite{gpry_1, gpry_2, gpry_code} to construct such a surrogate model.
In this subsection, we adopt the notation most commonly used in the context of Gaussian processes where $\mathbf{x}$ refers to a vector in the sampling space (equivalent to $\theta$ above) and $y$ is the value of the target function.
\gpry iteratively proposes points $\mathbf{x}$ in parameter space at locations where the expected gain in information about the posterior is maximized. With them, at every iteration, it builds an approximation of the posterior log-density function $\log p(\mathbf{x}|\mathcal{D})$ under some data $\mathcal{D}$ as the mean of a Gaussian process conditioned on the current set of \textit{training} samples $(\mathbf{X}, \mathbf{y})$, where $\mathbf{X}=\{\mathbf{x}^{(i=1,\ldots)}\}$ and $\mathbf{y}=\{\log p(\mathbf{x}^{(i=1,\ldots)})\}$:
\begin{equation}\label{eq:gp_description}
    \log p(\mathbf{x}|\mathcal{D})\sim \mathcal{GP}(0, k(\mathbf{x}, \mathbf{x}')|\mathbf{X}, \mathbf{y})\,.
\end{equation}
Here \(k(\mathbf{x}, \mathbf{x}')\) represents the covariance function, for which \gpry uses a $d$-dimensional inverse-squared Radial Basis Function (RBF) kernel allowing for a different length-scale in each dimension of the sampled parameter space:
\begin{align}
    k(\mathbf{x}, \mathbf{x}') = C^2 \prod_{i=1}^d \exp\left(\frac{(x_i-x_i')^2}{2 l_i^2}\right)\,,\label{eq:gp_cov_function}
\end{align}
with $C$ and $l$ representing, respectively, the output and length scales of the Gaussian process. The null mean of the Gaussian process prior in \cref{eq:gp_description} applies to a transformed set of $\mathbf{y}$ log-posterior values so that they have null mean and unit standard deviation.
Hereon we drop the explicit dependence on the training data $\mathbf{X}, \mathbf{y}$.

The mean of the conditioned Gaussian process, with which we approximate the log-posterior density, is computed as
\begin{equation}\label{eq:gp_mean}
    \mu(\mathbf{x}_*) = \mathbf{k}_*^T (\mathbf{K} + \sigma_n^2 \mathbf{I})^{-1} \mathbf{y}\,,
\end{equation}
where $(\mathbf{k}_*)_i = k(\mathbf{x}^{(i)}, \mathbf{x}_*)$, $(\mathbf{K})_{ij} = k(\mathbf{x}^{(i)}, \mathbf{x}^{(j)})$, and $\sigma_n^2$ is an estimate of the numerical uncertainty of log-posterior values.
The standard deviation of the conditioned Gaussian process, used in the acquisition function defined below, is $\sigma(\mathbf{x})=\sqrt{\mathrm{diag(\Sigma(\mathbf{x}))}}$, where
\begin{equation}\label{eq:gp_cov}
    \Sigma(\mathbf{x}_*) = k(\mathbf{x}_*, \mathbf{x}_*) - \mathbf{k}_*^T (\mathbf{K} + \sigma_n^2 \mathbf{I})^{-1} \mathbf{k}_*.
\end{equation}
The hyperparameters of the kernel, i.e., its output and length scales, hereon denoted collectively as $\mathbf{\Lambda}$, are determined by maximizing their marginalized likelihood \cite{gpml}. The mean and standard deviation of a Gaussian process conditioned on a set of training samples can be seen in the upper row of \cref{fig:likelihood-approximation_alt}.

Optimizing the kernel hyperparameters $\mathbf{\Lambda}$ eventually dominates the overhead of the algorithm, as it requires multiple kernel matrix inversions that scale as $\mathcal{O}(N^3)$, with $N$ being the number of training samples.
In order to mitigate this, we only perform a full re-fit of the hyperparameters at every few iterations of the algorithm (see \cref{app:gpry}). 
In general, overhead costs start making \gpry an impractical approach for dimensionalities larger than a few tens, depending on the cost of the likelihood.
The number of training samples, which drives the overhead costs, needed for accurate posterior reconstruction depends on the dimensionality of the problem. In exchange for this overhead, \gpry reduces the number of posterior evaluations required with respect to traditional samplers by a factor of $\mathcal{O}(10^2)$. Therefore, \gpry's advantage in performance increases for low dimensions and large costs per posterior evaluation. An approximate rule of thumb is that \gpry is faster for dimensionalities lower than a few tens when the posterior evaluation time is $\mathcal{O}(1\,\mathrm{s})$ or higher.

As a further refinement of the surrogate model, we multiply the GPR by a Support Vector Machine (SVM) classifier, of comparatively negligible computational cost, trained both on the evaluations used for the GPR, and those rejected because their log-posterior density is either negative infinity or very low with respect to the best training point.
This SVM is used to partition the parameter space into regions in which the true log-posterior is expected to return a finite, or negative infinity value; the latter is used as an exclusion region where future candidates are automatically rejected without evaluating their true log-posterior.

To enable active sampling, we introduce an acquisition function, denoted as \(a(\mathbf{x})\), which guides the sampling process by quantifying the expected utility of sampling the true posterior at each point in the parameter space:
\begin{equation}\label{eq:acquisition_function}
    a(\mathbf{x}) = \exp\left(2\zeta\cdot\mu \left(\mathbf{x}\right)\right) \left(\sigma(\mathbf{x}) - \sigma_n\right)\,,
\end{equation}
$\zeta$ is a scaling factor that balances exploration and exploitation.
The learning efficiency is maximized when this scaling factor is made dimensionality-dependent, increasingly encouraging exploration for larger dimensionalities: $\zeta=d^{-c}$, with $c>0$ \cite{gpry_1}.
In this paper, we empirically set this scaling factor to $\zeta=d^{-0.65}$, promoting exploitation slightly more than the value derived in \cite{gpry_1} for Gaussian distributions. The effect of evaluating sequentially at the optimum of the acquisition function can be seen in \cref{fig:likelihood-approximation_alt}.

The acquisition function is optimized through the NORA active sampling strategy described in \cite{gpry_2}: we draw MC samples from the mean $\mu(\mathbf{x})$ of the GPR using a Nested Sampler (NS), in our study \code{PolyChord} \cite{Handley:2015fda,2015MNRAS.453.4384H}. The acquisition function is then evaluated at the resulting NS samples, and its value is used to produce a pool of candidate points. This pool is ranked using the Kriging believer \cite{parallel_kriging_believer_2} prescription so that the $n$-th point is assigned a conditioned acquisition function value assuming a true posterior evaluation at the $n-1$ points above it. The optimal batch size is approximately equal to $d$ \cite{gpry_1}, making the \gpry algorithm efficiently parallelizable up to $d$ processes using MPI.

The use of a NS at the acquisition step, that explores the full surrogate posterior (as opposed to directly maximizing the acquisition function), makes it easier for \gpry to map a multimodal posterior, such as those expected in the sky localization parameters for low-latency signals, as demonstrated in \cite{gpry_2}. This ability can be further boosted by making the acquisition function more exploratory (lower scaling factor $\zeta$ in \cref{eq:acquisition_function}), and the exploration of the posterior more thorough (larger number of \textit{live points} of the NS).

At the end of every iteration, convergence is checked and considered reached as soon as one of two criteria is fulfilled at least twice consecutively: the value of the likelihood at the new proposed sampling locations is close enough to their GPR-predicted value (see \cite{gpry_1} for clarification), or the Gaussian-approximated Kullback-Leibler divergence\footnote{I.e., the Kullback-Leibler divergence, see \cref{eq:kl_continuous}, when distributions are approximated as multivariate Gaussians defined by their respective empirical means and covariance matrices.} between consecutive NORA NS runs is small enough.

\begin{figure}[ht]
  \includegraphics[width=0.8\columnwidth]{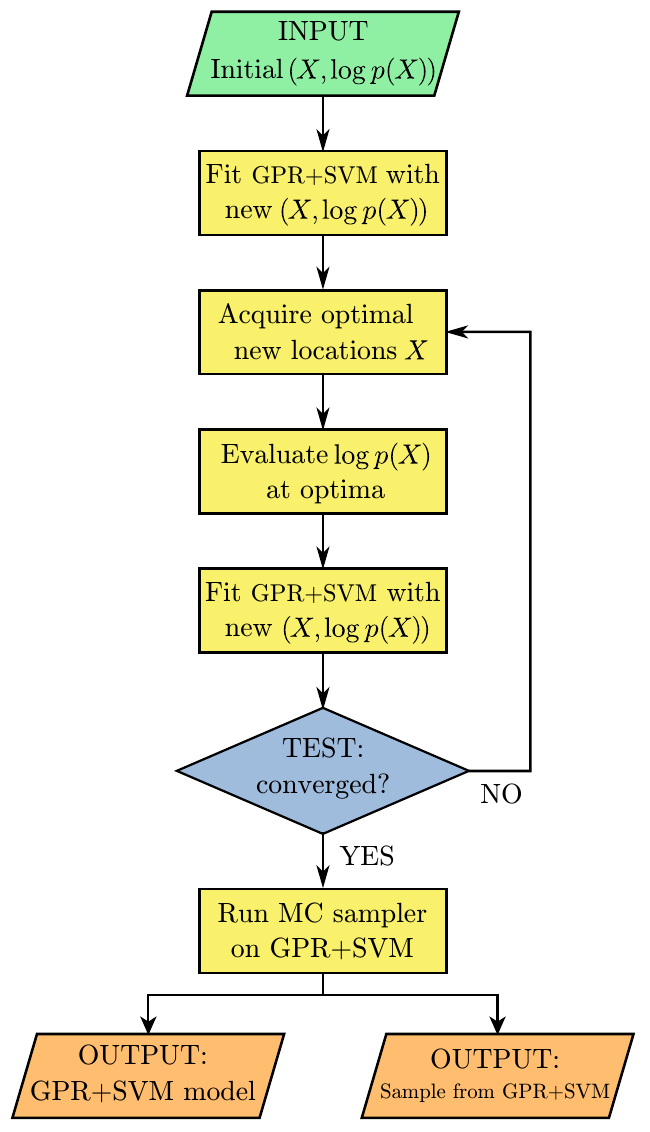}
  \caption{
    Simplified flow chart of the \gpry algorithm. Looking at \cref{fig:likelihood-approximation_alt}, the GPR at the top of its first column presents the initial stage, where the GPR has been fit to an initial set of two samples. The main loop (\textit{``acquire$\rightarrow$evaluate$\rightarrow$fit'')} corresponds sequentially to finding the location of the maximum of the acquisition function in the bottom row (dotted vertical line), evaluating the log-posterior there, and fitting the GPR to obtain the new model at the top of the following column. }
  \label{fig:gpry_flow}
\end{figure}

After convergence of the Bayesian optimization loop has been reached, MCMC samples of the surrogate model are generated. This typically only takes a few seconds since the evaluation of the surrogate model is very fast at $\mathcal{O}(10^{-5}\,\mathrm{s})$. To do so, we use \cobaya's implementation of the MCMC sampler of \cosmomc \cite{Lewis:2002ah, Lewis:2013hha}.\footnote{Notice that any other sampler, including those that require gradients, could be used without changing our conclusions.}
A flow chart of the algorithm is shown in \cref{fig:gpry_flow}.

\subsection{\label{sec:inf}Inference strategy and methodology for validating the results}

\gpry adopts default values for the parameters controlling the some of the aspects of the algorithm mentioned above, based on test runs on typical scenarios \cite{gpry_1,gpry_2}.
The peculiarities of the problem at hand motivate changing some of these defaults in each case, as detailed in \cref{tab:gpry_settings} in \cref{app:gpry}, and summarized below:

\begin{itemize}

\item For all three sources, especially for the \sbhb and \smbhb, the log-likelihood presents significant numerical noise with respect to small changes in the waveform parameters. If not correctly accounted for, \gpry interprets this sizable noise contribution as physically meaningful, which may lead to overfitting. We alleviate this problem by choosing large values of the expected noise scale $\sigma_n$ in \cref{eq:gp_cov}. Away from the mode, for very low likelihood values, the numerical noise dominates, so we raise the SVM classifier cutoff to exclude low-valued regions from the GPR.

\item For the sources with the slowest likelihood, the \sbhb and especially the \smbhb, it makes sense to increase the overhead of the algorithm in exchange for reducing the number of necessary true posterior evaluations for convergence. Hence, we increase the frequency and the number of restarts for the GPR hyperparameters optimization. Similarly, we update the set of NS samples from mean GPR more often, and, for the \smbhb, reduce the number of Kriging steps.

\item Since the set of initial points for the \sbhb and \smbhb is very informative (see \cref{sec:searches}), it is advantageous to define a \textit{trust region} around the current training set restricting the area where new evaluations are proposed. This region is the minimal hyper-rectangle containing training samples with posterior density above some cutoff with respect to the best one.

\end{itemize}

For each source type, we consider a high-SNR source signal as a noiseless LISA data stream, then inject it in multiple simulated noise realizations
The three easier noiseless inference  problems are used for consistency checks (e.g., robustness with respect to initialization) and are not presented below.

In order to benchmark \gpry's performance, both in terms of computational cost and inference accuracy, we pair every \gpry run in each noise realization with a similar run with the machine-learning-enhanced nested sampler \nessai, which has proven to be an efficient and reliable sampler in the context of GW data analysis~\cite{nessai,Williams:2021qyt,Williams:2023ppp}.

We perform two tests on the two sets of runs.
The first focuses on the accuracy of the full pipeline, from signal and noise generation to MC sampling.
In literature, this test is often referred to as a $pp$-plot~\cite{CookGelmanRubin,ppplots}.
For a given sampler choice and source category, we perform $N$ inference runs on independent noise realizations, and compute the empirical quantiles $\left\lbrace q_i\right\rbrace_i^N$ corresponding to the injected parameters for the inferred posterior.
In the limit $N\rightarrow\infty$ the cumulative distribution function of quantiles across runs is theoretically expected to approach that of the uniform distribution over the unit interval.
Deviations from the asymptotic distribution due to finite $N$ can be estimated numerically, and confidence intervals constructed accordingly.
We present results of this test across source categories in \cref{fig:WD_pp,fig:SOBBH_pp,fig:SMBBH_pp}, respectively.

In the second test, we focus instead on a direct comparison between posteriors obtained through inference with \nessai and \gpry in paired runs on the same noise realizations.
To do so, we evaluate the Jensen-Shannon (JS) divergence $\djs$ between each \nessai posterior distribution $P$ and the \gpry surrogate model $P_{\scriptscriptstyle \mathrm{GP}}$ over the parameter space \cite{61115}
\begin{equation}
  \djs(P||P_{\scriptscriptstyle \mathrm{GP}}) = \frac{1}{2}\left(\dkl(P||M) + \dkl(P_{\scriptscriptstyle \mathrm{GP}}||M)\right)\,,\hspace{-15pt}\label{eq:DJS}
\end{equation}
where $M = \frac{1}{2}(P+P_{\scriptscriptstyle \mathrm{GP}})$ is the mixture distribution of $P$ and $P_{\scriptscriptstyle \mathrm{GP}}$. The Kullback-Leibler (KL) divergence between two continuous probability distributions $P$, $M$ with densities $p(x)$, $m(x)$ is defined as
\begin{align}\label{eq:kl_continuous}
    D_{\scriptscriptstyle\mathrm{KL}}(P||M) = \int p(x)\log\left(\frac{p(x)}{m(x)}\right)\, \mathrm{d}x \ .
\end{align}
In practice, we compute the KL divergence as a Monte Carlo sum of the samples from \gpry and \nessai.
In this paper, we use natural logarithms for the divergence calculations.
The JS-divergence $\djs(P||Q)$ approaches zero if and only if $P$ and $Q$ describe the same distribution and is upper-bounded by $\log 2$. For inference purposes, values of $\djs\lesssim0.05$ would make \gpry as accurate as traditional samplers, whereas values up to $\djs=0.1$ could be considered precise enough, given the large computational trade-off.
We show the distribution of $\djs$ for different source categories in \cref{fig:WD_js,fig:SOBBH_js,fig:SMBBH_js}, respectively.

Following the formulas in \cref{sec:approx_djs}, for the dimensionality of our problems $\djs=0.05$ ($\djs=0.1$) would translate into a mean deviation in each parameter of $\approx 0.08\sigma$ ($\approx 0.11\sigma$) if assuming similar covariances, or alternatively a misestimation of the error of $\sim 15\%$ ($\sim 25\%$) if assuming similar means.

\section{\label{sec:results}Results}

\subsection{Double white dwarf system\label{sec:result_dwd}}

For a single injection of a \dwd system, the waveform and subsequent likelihood computations are fast ($\sim 10^{-3}\,\mathrm{s}$). Hence, we do not expect significant savings in wall clock computation time between \gpry and \nessai. We therefore use it to test the \gpry algorithm and gain some insight on its reliability.

\begin{table}[ht]
  \begin{tabular}{|l|c|l|}
    \hline
    \textbf{Parameter} & \textbf{Symbol} & \textbf{Value} \\\hline\hline
    Ecliptic longitude  & $\lambda$    & $2.0 \,{\rm rad}$ \\\hline
    Ecliptic sine-latitude & $\sin\beta$ & $0.479$ \\ \hline
    Amplitude & $A$                    & $2\cdot 10^{-23}$ \\\hline
    Frequency & $f$                    & $0.00377 \,{\rm Hz}$ \\ \hline
    Frequency derivative & $\dot{f}$   & $2\times 10^{-18} \,{\rm Hz^2}$ \\ \hline
    Cosine-inclination & $\cos \iota$  & $0.4$ \\ \hline
    Left phase & $\phi_L$              & $1.3 \,{\rm rad}$ \\ \hline
    Right phase & $\phi_R$             & $1.5 \,{\rm rad}$ \\ \hline
    \hline
    \multicolumn{2}{|l|}{\textbf{SNR}} & $23.64$ \\\hline
  \end{tabular}
  \caption{\label{tab:WD_params}
    Injected values for the sampled parameters of the \dwd system and total source ${\rm SNR}$.
  }
\end{table}

\begin{figure}[ht]
  \subfloat[\label{fig:WD_pp}]{
    \includegraphics[width=0.95\columnwidth]{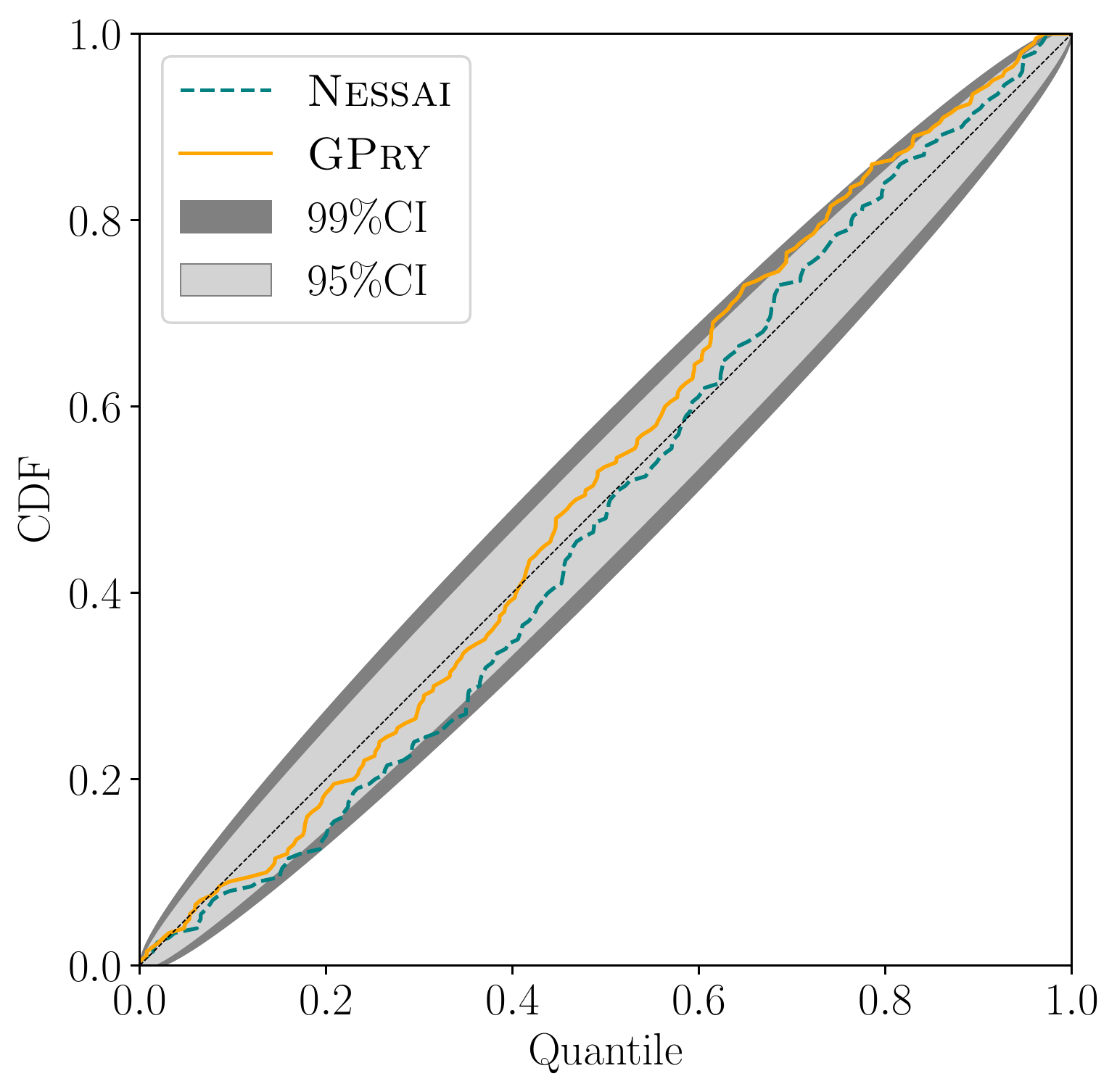}
  }\hfill 
  \subfloat[\label{fig:WD_js}]{
    \includegraphics[width=0.95\columnwidth]{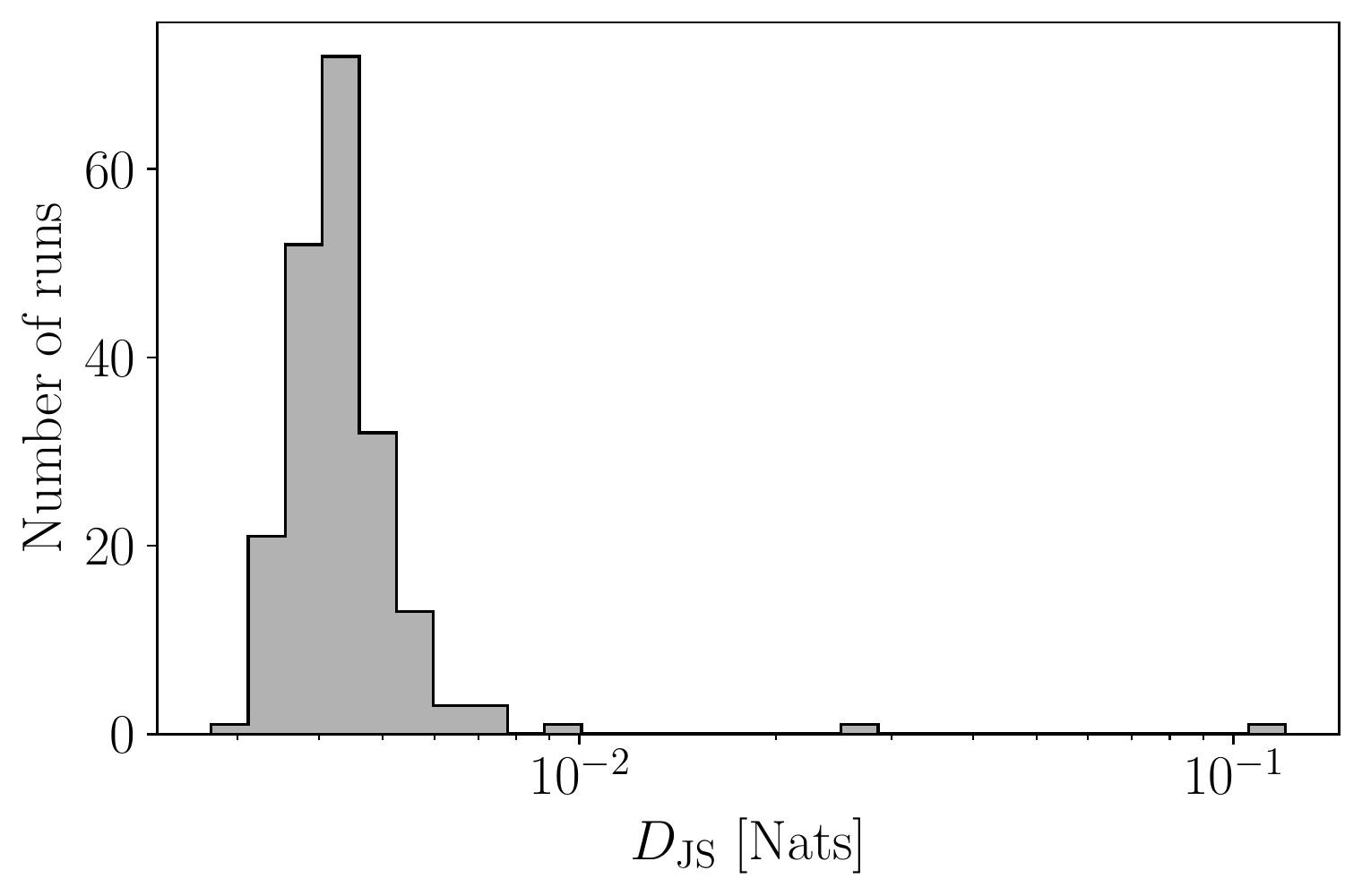}
  }
\caption{PP plot \textbf{(a)} and Jensen-Shannon divergence \textbf{(b)} for 200 \dwd runs with different noise realizations. \nessai and \gpry show comparable accuracy in the former, consistent at $99\%$ confidence (dark gray shaded area) with the theoretical prediction (dotted black line) across all runs, and at $95\%$ confidence (light gray shaded area) for the largest majority of them.
Relatively to \nessai, \gpry reconstructs the posterior shape reliably with only one run exceeding the target of $\djs=0.05$.}
\end{figure}

The injected parameters for the benchmark source are shown in \cref{tab:WD_params}. We draw 200 noise realizations according to our model in \cref{sec:data}.
For this source all parameters are constrained and the posterior distribution exhibits a single, localized nearly-Gaussian mode.
For each noise realization, we perform separate inference runs with \gpry and \nessai, with the \nessai runs performed with 2000 live points.
We then generate a PP plot comparing the performance of both algorithms (see \cref{fig:WD_pp}), and find a similar accuracy for the reconstruction.
Furthermore, we compute the JS divergence, $\djs$, between \gpry and \nessai for each noise realization, and show its histogram in \cref{fig:WD_js}.
This comparison shows that both samplers are in excellent agreement for all but one noise realization.
In \cref{fig:WD_corner_noisy} we show a corner plot overlaying the posterior contours obtained by \gpry and \nessai for the realization corresponding to the median $\djs$.
There we can observe the clear agreement between the two approaches, achieved with $1/300$ fewer likelihood evaluations by \gpry compared to \nessai.
The locations of the \gpry evaluations can be seen in the upper triangle of \cref{fig:WD_corner_noisy}.

In \cref{fig:WD_corner_noisy_failed} we show a corner plot and the posterior contours obtained by \gpry and \nessai corresponding to the highest $\djs=0.12$. Although the mode has been found by \gpry in this example, it remains underexplored. Tightening the convergence criterion would eliminate this problem in exchange for higher computational costs, but maintaining the two-orders-of-magnitude difference in the number of likelihood evaluations with respect to \nessai. Only one of the 200 runs performed shows this behavior with $\djs>0.05$ which leads us to conclude that the precision and accuracy of \gpry is sufficient in this context.

\subsection{\label{sec:results_sobbh}Stellar origin binary black holes}

For one injection of a \sbhb system, the cost of a single evaluation of the inference pipeline (waveform and likelihood calculations) is $\sim 10^{-1}\,\mathrm{s}$, which is significantly higher than for \dwds.
Thus, the savings here are potentially higher, which makes \gpry a worthy approach.

\begin{table}[ht]
  \begin{tabular}{|l|c|l|}
    \hline
    \textbf{Parameter} & \textbf{Symbol} & \textbf{Value} \\\hline\hline
    Redshifted chirp mass & ${\cal M}_c$ & $ 48.618 \, \msun$ \\\hline
    Reduced mass-ratio & $\delta \mu$ & $ 0.5 $ \\\hline
    Ecliptic longitude & $\lambda$ & $ 0.19 \, {\rm rad}$ \\\hline
    Ecliptic sine-latitude & $\sin\beta$ & $ 0.82 \, {\rm rad}$ \\\hline
    Initial orbital frequency & $f_0$ & $ 1.87 \, {\rm mHz}$ \\\hline
    Left phase (fixed) & $\phi_L$ & $ 0.97 \, {\rm rad}$ \\\hline
    Right phase (fixed) & $\phi_R$ & $ 1.76 \, {\rm rad}$ \\\hline
    Left square-root amplitude & $\sqrt{A_L}$ & $12.57\cdot 10^{-5}$ \\\hline
    Right square-root amplitude & $\sqrt{A_R}$ & $1.13\cdot 10^{-5}$\\\hline
    Dimensionless spin & $\chi_1$ & $ 0.223$ \\\hline
    Dimensionless spin & $\chi_2$ & $ 0.262$ \\\hline\hline
    \multicolumn{2}{|l|}{\textbf{SNR}} & $ 16.79$\\\hline
  \end{tabular}
  \caption{\label{tab:SOBBH_params}
    Injected values for the parameters of the \sbhb system, and total source ${\rm SNR}$. All parameters are sampled except for the phases, for which the method described in \cref{sec:searches} failed to provide reliable estimates. The detector-frame individual masses are $m_1=99.55\,\msun$ and $m_2=33.18\,\msun$.
    }
\end{table}

The benchmark source's injected parameters are shown in \cref{tab:SOBBH_params}. Unfortunately, as the semi-coherent search presented in \cref{sec:searches} does not provide us with a reliable estimate of the phases, sampling these proves to be difficult with \gpry. This is further complicated by the periodic nature of these parameters. We therefore fix the values to the injected ones.
Contrary to the \dwd case, the resulting posterior is highly non-Gaussian: it is heavy-tailed and has a large curving degeneracy.
As discussed in \cite{gpry_1}, exploring the full posterior in a reasonable amount of time poses a challenge to \gpry.

\begin{figure}[t]
  \subfloat[\label{fig:SOBBH_pp}]{
    \includegraphics[width=0.95\columnwidth]{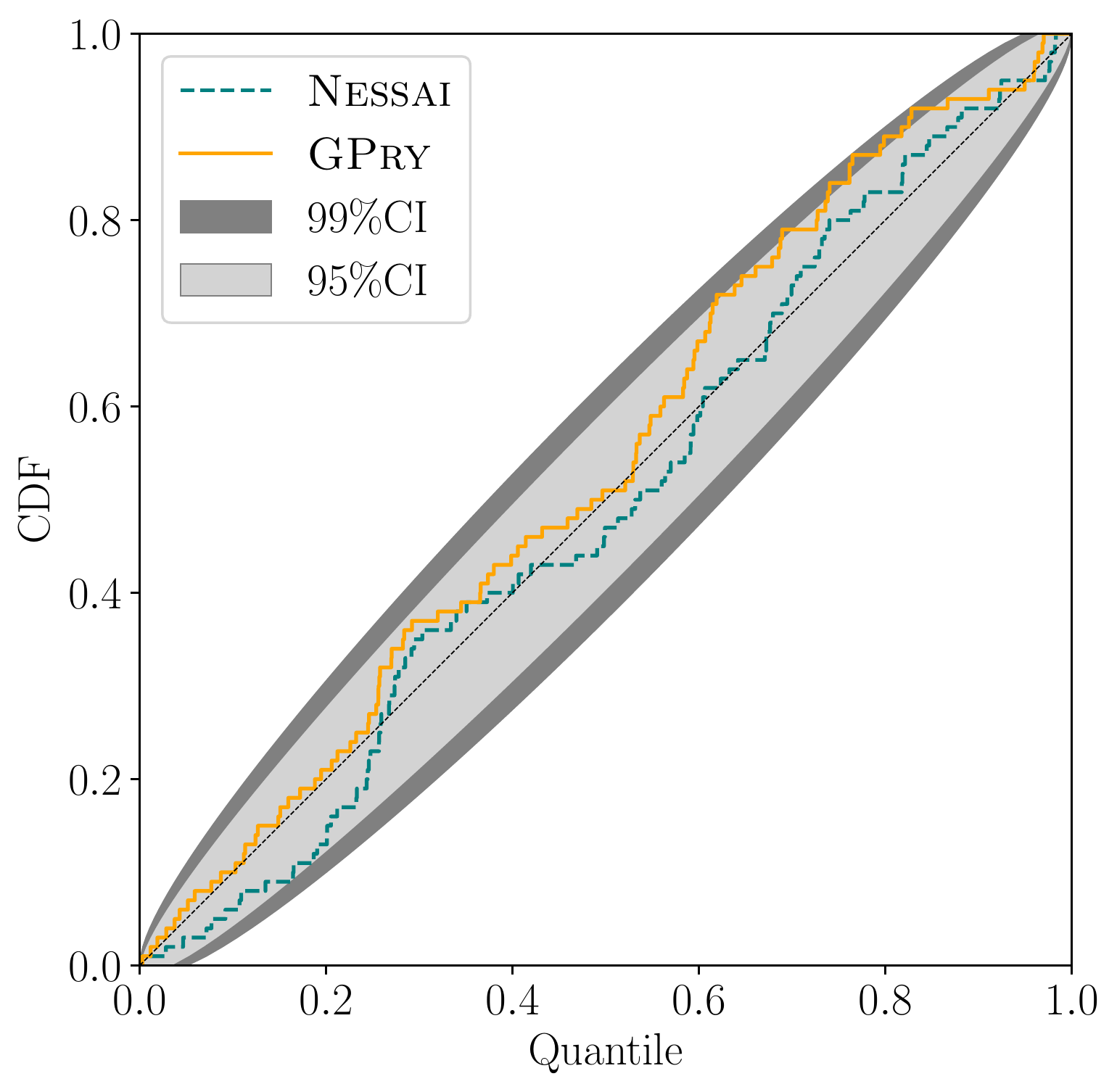}
  }\hfill
  \subfloat[\label{fig:SOBBH_js}]{
    \includegraphics[width=0.95\columnwidth]{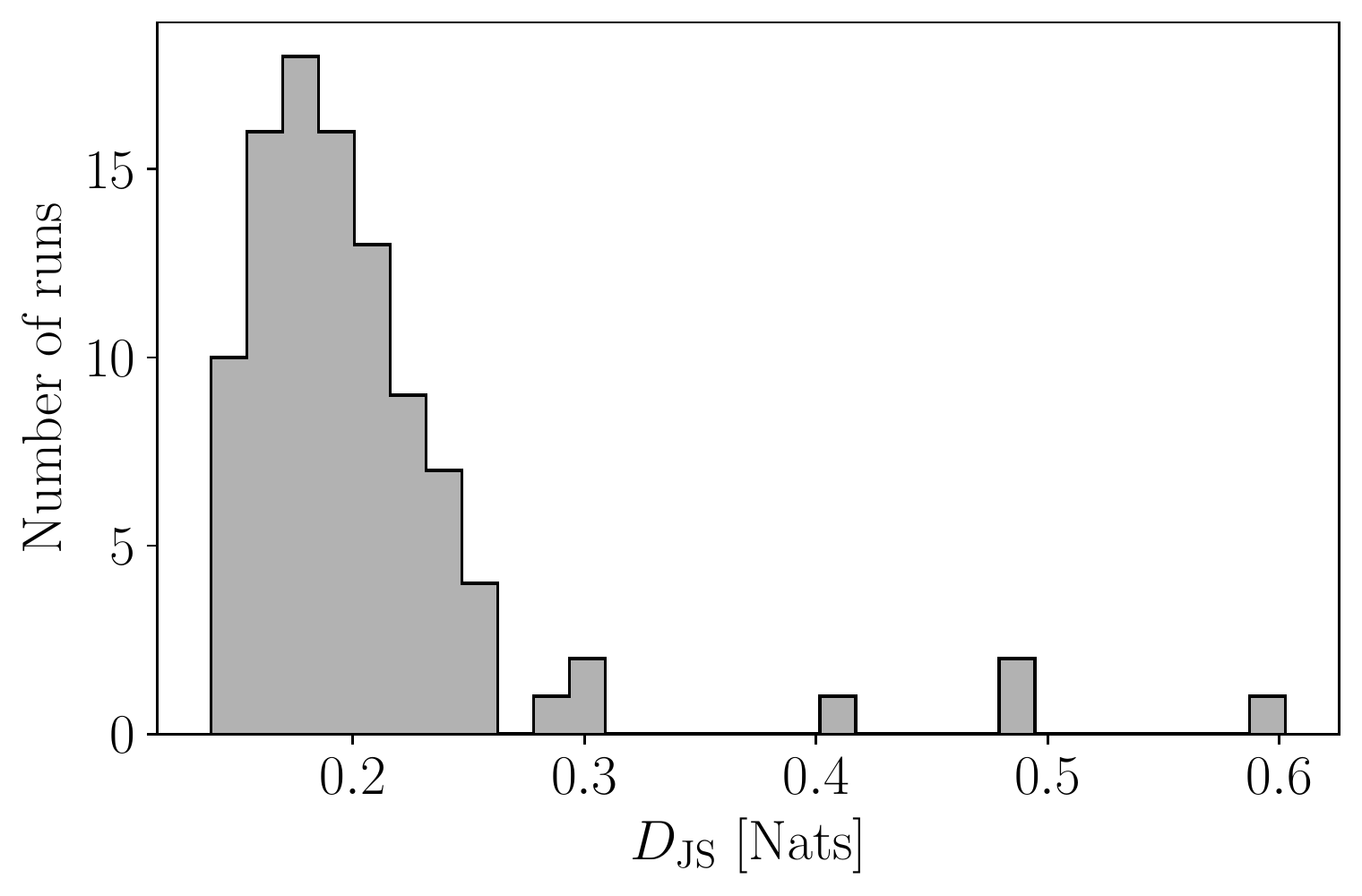}
  }
  \caption{
    PP plot \textbf{(a)} and Jensen-Shannon divergence \textbf{(b)} for 100 \sbhb runs with different noise realizations. While \nessai and \gpry show comparable accuracy in the former, consistent at $99\%\,\mathrm{CL}$ (dark shaded region) with the theoretical prediction (black dotted line), the distribution of $\djs$ that is entirely above the target value of $0.05$ indicates insufficient characterization of the posterior mode. Indeed, \cref{fig:SOBBH_corner_noisy} shows that \gpry underestimates the tails, especially in the $\mathcal{M}_c,\delta\mu$ direction which leads to the large discrepancy.}.
  \label{fig:SOBBH_pp_js}
\end{figure}

We generate 100 noise realizations and perform inference runs for each of them with \gpry and \nessai, with the \nessai runs performed with 2000 live points.
We then generate both a PP plot (see \cref{fig:SOBBH_pp}), and compute the JS divergence for each pair of runs, whose histogram is shown in \cref{fig:SOBBH_js}.
From the PP plot, it is clear that \gpry performs worse than \nessai, even if both samplers show reasonably good performance.
This is reflected in the higher JS divergences between \nessai and \gpry (see \cref{fig:SOBBH_js}), localized mostly in the $[0.1, 0.25]$ interval, with a few $\djs\gtrsim0.3$ outliers.

The effect of the $\djs\sim0.2$ divergence is illustrated in \cref{fig:SOBBH_corner_noisy}, which shows the result of the median $\djs$ run with \gpry and \nessai.
As we can see, although the resulting mode for \gpry is localized correctly towards the injected value, it fails to explore a fraction of the posterior corresponding to the large-values tail of the $(\mathcal{M}_c,\delta\mu)$ degeneracy. The handful of cases with higher $\djs$ (up to $0.6$) present the same sort of effect, and small ($<1\sigma$) biases for other parameters.

Possible mitigation strategies include fine-tuning of the \gpry hyperparameters, to increase the chance that it fits this particular problem better (e.g.\ that it does not converge prematurely), as well as the use of alternative parameterizations whose posterior would not present these strong non-Gaussian features.
We leave this endeavor for future work. It must be remarked that this difficulty also affects \nessai, whose precision we had to increase in order to map this posterior correctly, so that $\mathcal{O}(10^2)$ more evaluations are needed than for the other two test sources (see \cref{fig:nsamples}).

As explained in \cref{sec:searches}, we are seeding the \sbhb runs discussed in this section with high-likelihood points from a noiseless Semi-Coherent PSO run. Since a noise realization introduces a bias in the inferred source parameters with respect to the noiseless case, we have investigated whether this under performance may be related to the use of a biased initial set of samples in the \gpry runs. To do this, we have performed 100 additional paired runs with a noiseless injection, but found the same under-exploration effect with a similar magnitude.

We find that our approach reliably recovers the expected central values, and therefore could be used for source subtraction or fast, preliminary analysis; it is however suboptimal for full statistical source characterization. There is ongoing development of \gpry addressing more robust inference in highly non-Gaussian distributions such as this \sbhb posterior.

\subsection{Supermassive binary black-hole}

We now focus on the injection of a single \smbhb in noisy LISA data. Here, the cost per evaluation of the inference pipeline is larger than a few seconds and therefore \gpry shows great potential: it could turn days- or weeks-long inference runs with \nessai into hours-long ones.

\begin{table}[ht]
\begin{tabular}{|l|c|l|}
\hline
\textbf{Parameter} & \textbf{Symbol} & \textbf{Value} \\
\hline
\hline
Redshifted chirp mass & ${\cal M}_c$ & $ 6.5744 \times 10^6 \, \msun$ \\\hline
Reduced mass-ratio & $\delta \mu$ & $ 0.12864 $ \\\hline
Luminosity distance & $d_L$ & $ 18.7 \, {\rm Gpc}  $ \\\hline
Ecliptic longitude & $\lambda$ & $ 2.15 \, {\rm rad}$ \\\hline
Ecliptic sine-latitude & $\sin\beta$ & $ -0.34 \, {\rm rad}$ \\\hline
cosine-inclination & $\cos\iota$ & $ 0.86 \, {\rm rad}$ \\\hline
Orbital phase & $\phi^{\rm orb}_0$ & $ 5.86 \, {\rm rad}$ \\\hline
Polarization & $\psi$ & $ -0.136 \, {\rm rad}$ \\\hline
Time to merger & $\tau_m$ & $ 4.138 \times 10^6\,\mathrm{s}\ (47.89 \, {\rm days})$ \\\hline
Dimensionless spin & $\chi_1$ & $ 0.9874$ \\\hline
Dimensionless spin & $\chi_2$ & $ 0.9876$ \\\hline\hline
\multicolumn{2}{|l|}{\textbf{SNR}} & $ 1944.8$ \\\hline
\end{tabular}
\caption{\label{tab:SMBBH_params}
  Injected values for the sampled parameters of the \smbhb system and total source ${\rm SNR}$.
  The detector-frame individual masses are $m_1=8.61 \times 10^6 \, \msun$ and $m_2=6.65 \times 10^6 \, \msun$.
  The reference frequency is $f_\mathrm{ref}=10^{-4}\,\mathrm{Hz}$.
  }
\end{table}

The injected parameters for the benchmark source are shown in \cref{tab:SMBBH_params}. For this high signal-to-noise case, the posterior is nearly Gaussian.

\begin{figure}[ht]
  \subfloat[\label{fig:SMBBH_pp}]{
    \includegraphics[width=0.955\columnwidth]{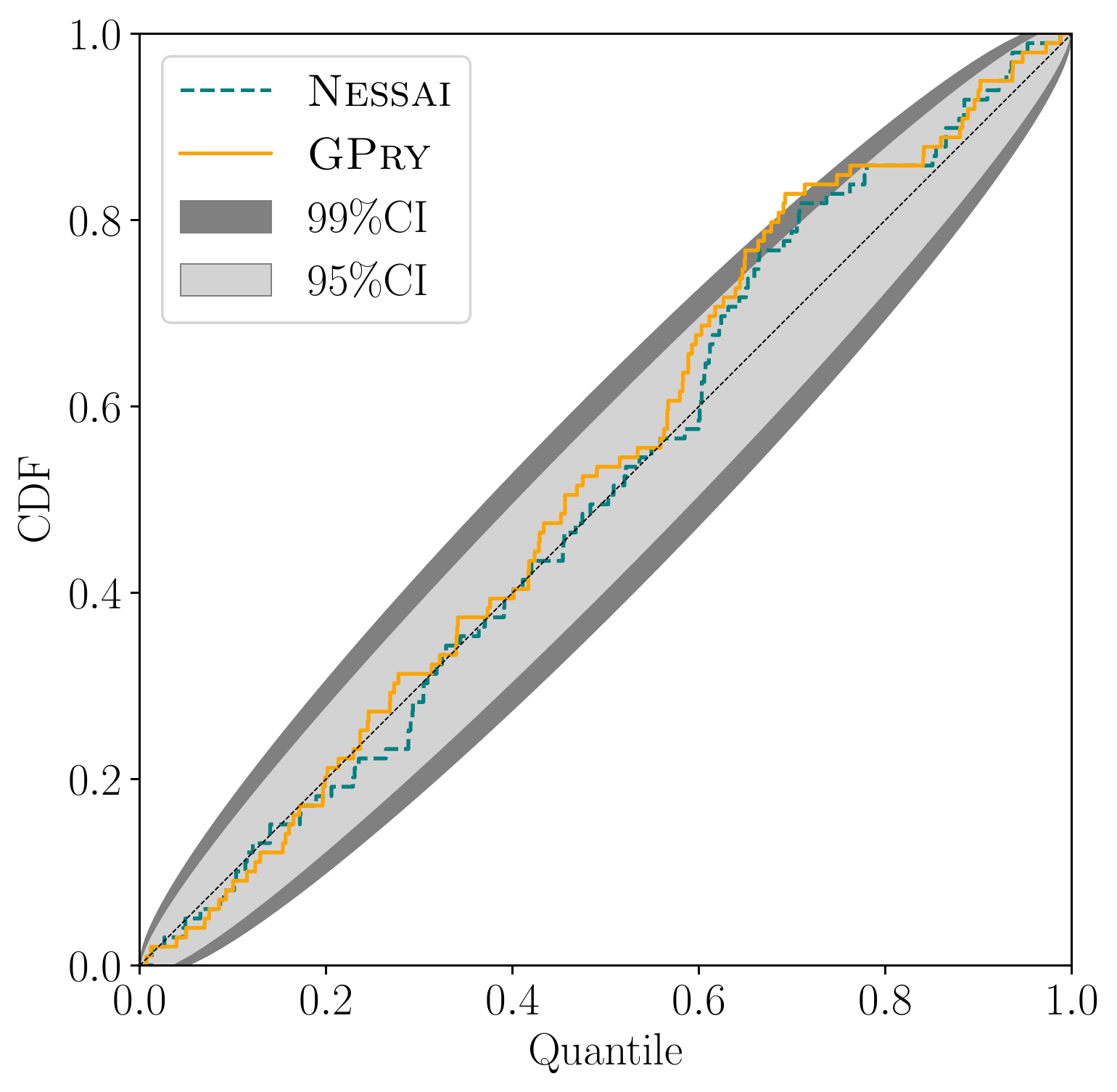}
  }\hfill 
  \subfloat[\label{fig:SMBBH_js}]{
    \includegraphics[width=0.95\columnwidth]{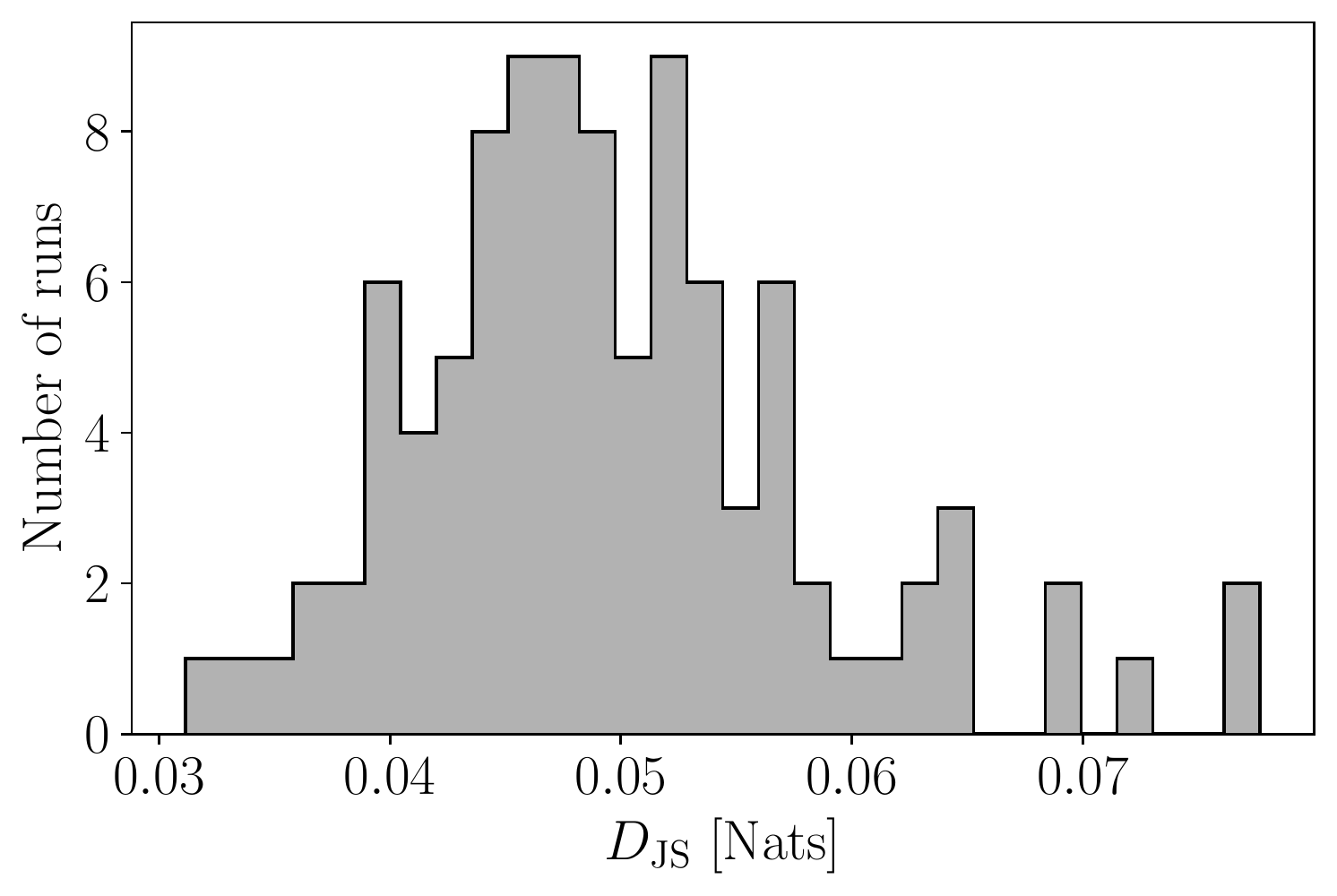}
  }
  \caption{
    PP plot \textbf{(a)} and Jensen-Shannon divergence \textbf{(b)} for 99 \smbhb noisy runs. The former shows that both \gpry and \nessai show comparable accuracy, consistent at $99\%$ confidence (dark gray shaded area) with the theoretical prediction (black dotted line).
    The distribution of $D_\mathrm{JS}$ clusters around our target value of $0.05$. This might partially be caused by \nessai running at low resolution, but also by \gpry occasionally under-exploring the tails of the posterior.}
  \label{fig:SMBBH_pp_js}
\end{figure}

In this case, we generate 100 noise realizations (of which one was discarded due to an HPC error) and perform inference runs with \gpry and \nessai.
The \nessai runs were performed with 500 live points, instead of 2000 as for the other sources, for reasons of limited computational capacity.
The resulting PP plot can be seen in \cref{fig:SMBBH_pp}, and the relative JS divergence for each pair of runs in \cref{fig:SMBBH_js}.
Both \gpry and \nessai show very good performance in the PP plot, and agree very well, with a median $\djs\approx0.05$ and no run with $\djs\ge0.1$. The runs with the median and highest $\djs$ are shown in \cref{fig:SMBBH_corner_noisy,fig:SMBBH_corner_noisy_notsogood}, respectively.
Therein we contrast the respective \gpry runs with two higher-resolution (2000 live points) \nessai runs performed to show finer contours for comparison.

For the \smbhb runs \gpry needs $n<10^3$ evaluations, which amounts to $\approx 30\%$ of the total computation time when the learning overhead is accounted for. In contrast, \nessai, despite being run with a significantly low resolution for reasons of time, performs $n\sim 10^5$ evaluations.

\subsection{Number of posterior evaluations and speedup}

\begin{figure*}[ht]
  \subfloat[\label{fig:nsamples}]{
    \includegraphics[width=0.99\columnwidth]{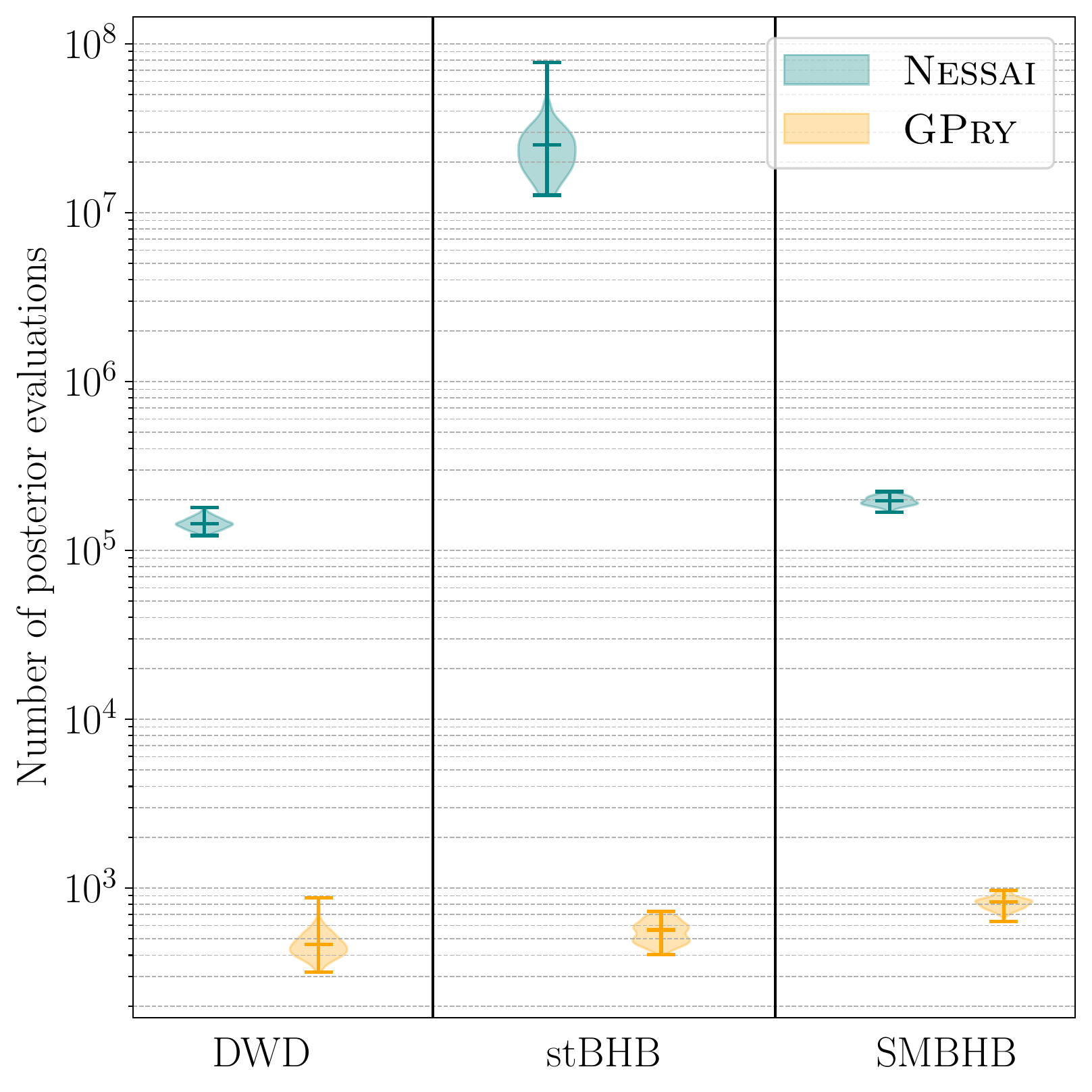}
  }
  \subfloat[\label{fig:sampling_times}]{
    \includegraphics[width=0.99\columnwidth]{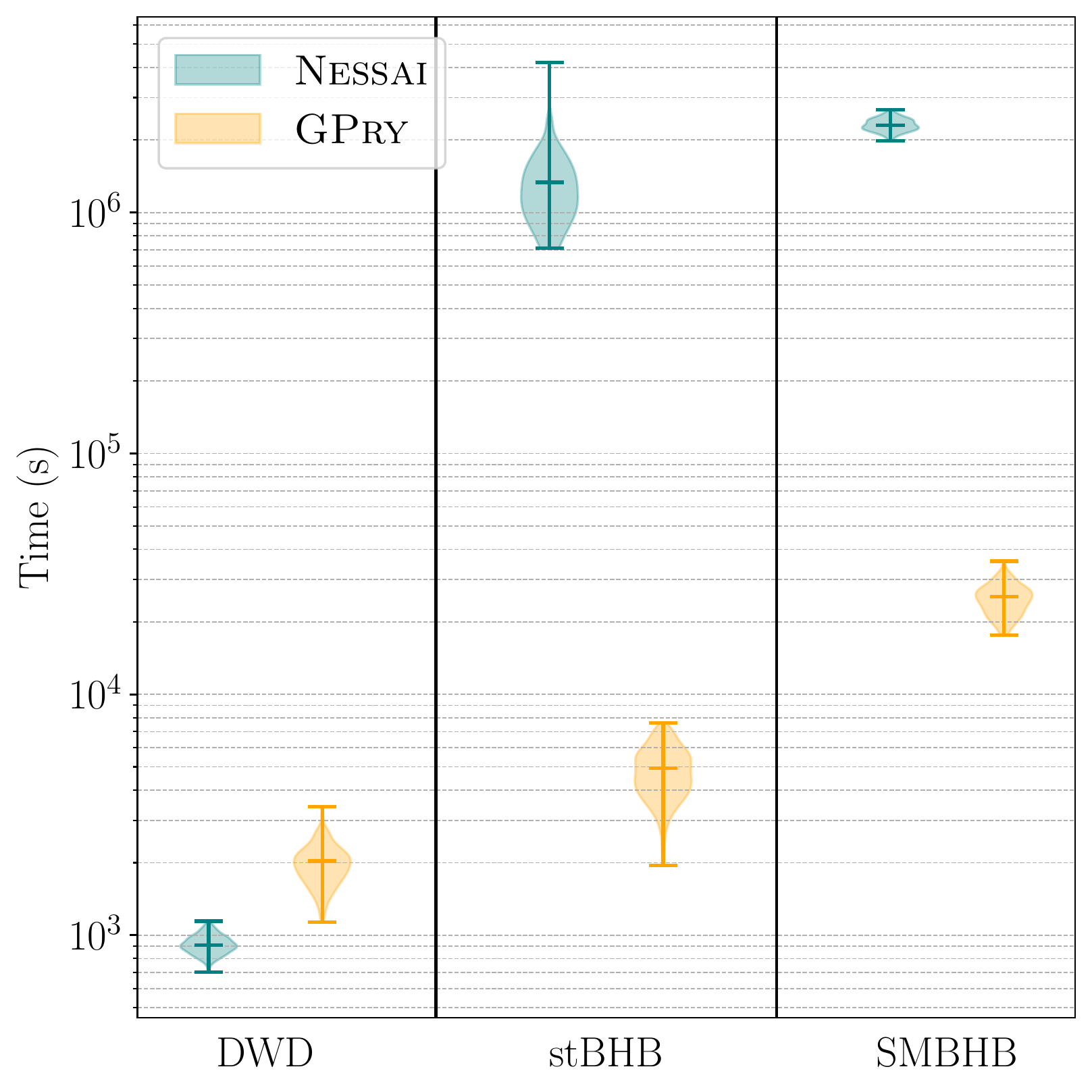}
  }
  \caption{
    Violin plots comparing \gpry (orange) and \nessai (teal) on each of the three test sources on noisy LISA data, according to \textbf{(a)} the total number of posterior evaluations needed, and \textbf{(b)} the hypothetical wall-clock time required for inference in a single-core setup (see text for a precise definition of these time estimates). The violins show the distribution the number of posterior samples and times, the minimum, median, and maximum values are marked with horizontal bars.
    }
  \label{fig:nsamples_times}
\end{figure*}

\gpry's main advantage compared to more traditional samplers is a drastic reduction in the number of posterior evaluations needed for inference, as shown in \cref{fig:nsamples}. It is clear that \gpry consistently performs $\mathcal{O}(10^2)-\mathcal{O}(10^3)$ fewer evaluations than \nessai to converge to the posterior mode. This, however, comes at the price of the relatively large amount of time required for the acquisition of new optimal sampling locations and fitting the GPR hyperparameters. The size of this overhead depends mainly on the dimensionality of the sampling space and, to a lesser extent, on the Gaussianity of the posterior. The dimensionality scaling of the overhead can be clearly observed in \cref{fig:gpry_overhead_vs_post} in \cref{app:gpry}. Ultimately, the potential speedup with respect to an alternative sampler depends on a combination of a slow-enough posterior and a reasonable dimensionality.

In \cref{fig:sampling_times} we show a comparison between the distribution of wall-clock times of the \gpry runs in this paper, and an optimistic (assuming no overhead) estimate for the paired \nessai runs. The quoted clock times are obtained by multiplying the number of their likelihood evaluations by their evaluation on the same hardware as for the \gpry runs.\footnote{The \nessai runs needed to be performed on a different platform due to limitations in our computing budget.} As we can see there, in the case of the \dwd source \gpry does not outperform \nessai, needing roughly twice the time on average, due to the very short computation time of the \dwd likelihood. However, for the \sbhbs and \smbhbs, where likelihood computations are more expensive, the speed up is highly significant, reducing the time for inference from $\sim 10^6$ core seconds (around 11 days) to $\sim 10^4$ core seconds (around 3 hours). Of course, both of these numbers can be reduced through parallel processing but the advantage would still be evident.

When taking the reliability and accuracy of the inference into account, it is clear that the biggest potential for speed up is currently in the inference of \smbhbs.

\section{\label{sec:conclusions}Conclusions}

We demonstrated that active sampling methods with Gaussian processes have the ability to produce accurate inference on individual injections of three different GW sources expected in the LISA band, \dwds, \sbhbs and \smbhbs, employing $\mathcal{O}(10^{-2})$ fewer evaluations of the GW signal likelihood than a state-of-the-art nested sampler, and with a significant speedup, going up to a $\mathcal{O}(10^{-2})$ wall-clock time reduction for likelihood evaluation times approaching $\mathcal{O}(1\,\mathrm{s})$ and above.
They do so with some, but little preconditioning, that can be provided by frequentist searches or other faster but less accurate approximate inference schemes.
Crucially, no expensive pretraining is required with these methods as would be in amortized approaches.

Using \gpry as an active learning framework, we found the advantages with respect to traditional Monte Carlo samplers to be problem-dependent:
\begin{itemize}
    \item Inference for \dwds can be provided quickly and robustly, but the fast-to-evaluate \dwd waveforms mean that the overhead of acquiring samples and fitting the Gaussian process outweigh the time saved by reducing the number of sampling steps. This in turn means that we report no savings in terms of wall-clock time. However, in the presence of gaps in the data, as expected in LISA, the computational cost of the likelihood will go up. In this case our approach could be competitive.

\item
Inferring the parameters for \sbhbs was possible with considerable time savings of 2 orders of magnitude. However, this comes at the cost of underestimating the tails of the distribution, though we retain the ability to reliably recover the central values. There is ongoing development of \gpry aimed at addressing this shortcoming.

\item
The best combination of speed-up and accuracy was achieved for the \smbhbs, whose likelihood is very slow to evaluate at $\mathcal{O}(1\,\mathrm{s})$. We report a speed-up of two orders of magnitude compared to nested sampling while retaining a comparable accuracy. This reduces the computational cost of the inference from $\sim 10^6$ core-seconds ($\sim 11$ core-days) to merely $\sim 10^4$ core-seconds ($\sim 3$ core-hours).

\end{itemize}

Integrating \gpry (or other active learning approaches) into the LISA Global Fit pipeline has the potential to yield a number of advantages.
On one hand rapid searches, e.g., based on template banks~\cite{2025PhRvD.111d3045C} or machine-learning approaches~\cite{2024PhRvD.110f2003H}, provide source parameter estimates for individual bright signals.
On the other hand, large and computationally intensive analyses will update such estimates over longer timescales, taking into account the superposition of multiple signals in the same datastream.
While the former is expected to use a limited amount of data, e.g. the premerger inspiral-only signal for SMBBHs, the latter will use all available data in high latency.

In addition to this, current Global Fit architectures under development, will fill the gap between the two with so-called deep-low-latency analyses: inference routines targeting low-dimensional, single-source parameter spaces, though flexible enough to condition on data knowledge acquired by the Global Fit.
Thanks to its active-learning approach, this is arguably the best venue to leverage \texttt{GPry}'s flexibility.
Global Fit conditional likelihoods are highly nontrivial to simulate in advance, hence being difficult to incorporate in the low-latency routines mentioned above. 
Acquiring an accurate reconstruction of the conditional posterior surface through a limited number of likelihood evaluations, \texttt{GPry} serves precisely this purpose.
This is advantage would be especially pronounced whenever new physics models are tested, where neither phenomenological waveforms may be suitable, or available pre-trained emulators may not be cost-effective.

\gpry can also be a very powerful tool for prototyping waveforms, theoretical models, and the inference pipeline with mock data. It requires no pretraining, and no noise to be present in the data and accounted for in the likelihood. This enables quick forecasting and testing without the need for dedicated computing infrastructure to be in place.

The resulting surrogate posterior can be stored as $~\mathrm{kB}$-sized object, a size much smaller than the data necessary to reproduce the inference problem, and can be upsampled at very low computational cost. As an analytic function, it can be easily used as a prior in subsequent searches or constraints.

In the future, we aim to go beyond the results of this paper in parallel with the ongoing development of \gpry, improving its accuracy in highly non-Gaussian and highly multimodal cases (e.g., extreme mass ratio inspirals), and its performance in larger dimensionalities such as inference problems with multiple sources.

\begin{acknowledgments}

The authors would like to thank A.~Klein, D.~Bandopadhyay, C.~Moore, and all the \balrog developers for useful discussions and insightful comments.
JE and GN acknowledge support from the ROMFORSK grant project no.~302640.
JE acknowledges support by the Spoke 3 (INAF) of the Italian Center for SuperComputing (ICSC), funded by the European Union - NextGenerationEU program, under the grant agreement N. C53C22000350006 (acronym Fab-HPCc).
RB acknowledges support from the ICSC National Research Center funded by NextGenerationEU, and the Italian Space Agency grant \emph{Phase A activity for LISA mission}, Agreement n.2017-29-H.0.
JT acknowledges financial support from the Supporting TAlent in ReSearch@University of Padova (STARS@UNIPD) for the project ``Constraining Cosmology and Astrophysics with Gravitational Waves, Cosmic Microwave Background and Large-Scale Structure cross-correlations'', and from a Ramón y Cajal contract by the Spanish Ministry for Science, Innovation and Universities with Ref.\ RYC2023-045660-I.
Computational resources were provided by University of Birmingham BlueBEAR High Performance Computing facility, by CINECA through EuroHPC Benchmark access call grant EUHPC-B03-24, by the Google Cloud Research Credits program with the award GCP19980904, and by the CloudVeneto initiative of the Università di Padova and the INFN -- Sezione di Padova.

\textbf{Software:}
We acknowledge usage of
\code{Mathematica}~\cite{Mathematica}
and of the following
\code{Python}~\cite{10.5555/1593511}
packages for modeling, analysis, post-processing, and production of results throughout:
\nessai~\cite{2021PhRvD.103j3006W},
\code{matplotlib}~\cite{2007CSE.....9...90H},
\code{numpy}~\cite{2020Natur.585..357H},
\code{scipy}~\cite{2020NatMe..17..261V},
\code{scikit-learn}~\cite{scikit-learn},
\cobaya~\cite{Torrado:2020dgo} and
\code{corner}~\cite{corner}.
\end{acknowledgments}
\bibliographystyle{apsrev4-1}
\bibliography{biblio}
\vfill
\clearpage
\onecolumngrid
\newpage
\appendix

\section{Approximate Jensen-Shannon divergence between multivariate Gaussians}\label{sec:approx_djs}

The KL divergence $\dkl$, defined in \cref{eq:kl_continuous}, has an analytical representation when computed between two $d$-dimensional multivariate Gaussians, $\mathcal{N}(\ve{\mu}_1,\ve{\Sigma_1})$ and $\mathcal{N}(\ve{\mu}_2,\ve{\Sigma_2})$, with means $\ve{\mu}_1$ and  $\ve{\mu}_2$ and covariance matrices $\ve{\Sigma}_1$ and $\ve{\Sigma}_2$. Conversely, an analytical representation for the JS divergence, defined in \cref{eq:DJS}, does not exist in this case. However, we can find an approximate expression if the mixture distribution $M$ (with mean $\boldsymbol{\mu}_M=\frac{1}{2}(\boldsymbol{\mu}_{1}+\boldsymbol{\mu}_{2})$) is a multivariate normal distribution with covariance $\boldsymbol{\Sigma}_M = \frac{1}{2}(\sigma_1^2+\sigma_2^2)\mathbb{I}$. The approximation holds if the multivariate Gaussians are sufficiently similar, i.e. $|\Delta\boldsymbol{\mu}|\equiv|\boldsymbol{\mu}_1-\boldsymbol{\mu}_2|\ll |\boldsymbol{\Sigma}_1 + \boldsymbol{\Sigma}_2|$ and $\sigma_1\approx \sigma_2$. In this case, the JS divergence reads 
\begin{align}
    \djs[\mathcal{N}(\boldsymbol{\mu}_{1}, \boldsymbol{\Sigma}_{1})||\mathcal{N}(\boldsymbol{\mu}_{2}, \boldsymbol{\Sigma}_{2})] \approx \frac{1}{4} \Delta\boldsymbol{\mu}^T(\boldsymbol{\Sigma}_1+ \boldsymbol{\Sigma}_2)^{-1}\Delta\boldsymbol{\mu} + \frac{1}{2}\log\left(\frac{|\boldsymbol{\Sigma}_1+ \boldsymbol{\Sigma}_2|}{\sqrt{|\boldsymbol{\Sigma}_1|| \boldsymbol{\Sigma}_2|}}\right)  - \frac{d}{2}\log 2\ .
\end{align}
For $\boldsymbol{\Sigma}_1 = \boldsymbol{\Sigma}_2 = \sigma^2\mathbb{I}$ this simplifies to
\begin{align}\label{eq:djs_gaussian_mu}
    \djs \approx  \frac{(\Delta\boldsymbol{\mu})^2}{8\sigma^2}\ ,
\end{align}
and for $|\Delta\ve{\mu}|=0$, $\boldsymbol{\Sigma}_1 = \sigma_1^2\mathbb{I},\ \boldsymbol{\Sigma}_2 = \sigma_2^2\mathbb{I}$ to
\begin{align}\label{eq:djs_gaussian_sigma}
    \djs \approx \frac{d}{2}\log\frac{\sigma_1^2+\sigma_2^2}{2\sigma_1\sigma_2}\ .
\end{align}

Therefore,  when only the mean of the distribution is misestimated,  $\djs$ is a function of the distance $\Delta\ve{\mu}$ in units of $\sigma$ (see \cref{eq:djs_gaussian_mu}). This is independent of the number of dimensions if only one parameter is misestimated whereas it is proportional to $d$  if the misestimation occurs for multiple parameters. On the other hand, if the mean is properly estimated but the spread of the distribution is not, then the result is proportional to $d$ whenever, on average, the spread is wrong by the same amount (see \cref{eq:djs_gaussian_sigma}).
We find that in less than 11 dimensions (the highest number of inferred parameters that we consider in this paper), the approximation holds up to $\djs\approx 0.1$.

\section{\label{app:gpry}Hyperparameters and overhead of \gpry}

\begin{figure}[ht]
    \centering
    \includegraphics[width=0.5\linewidth]{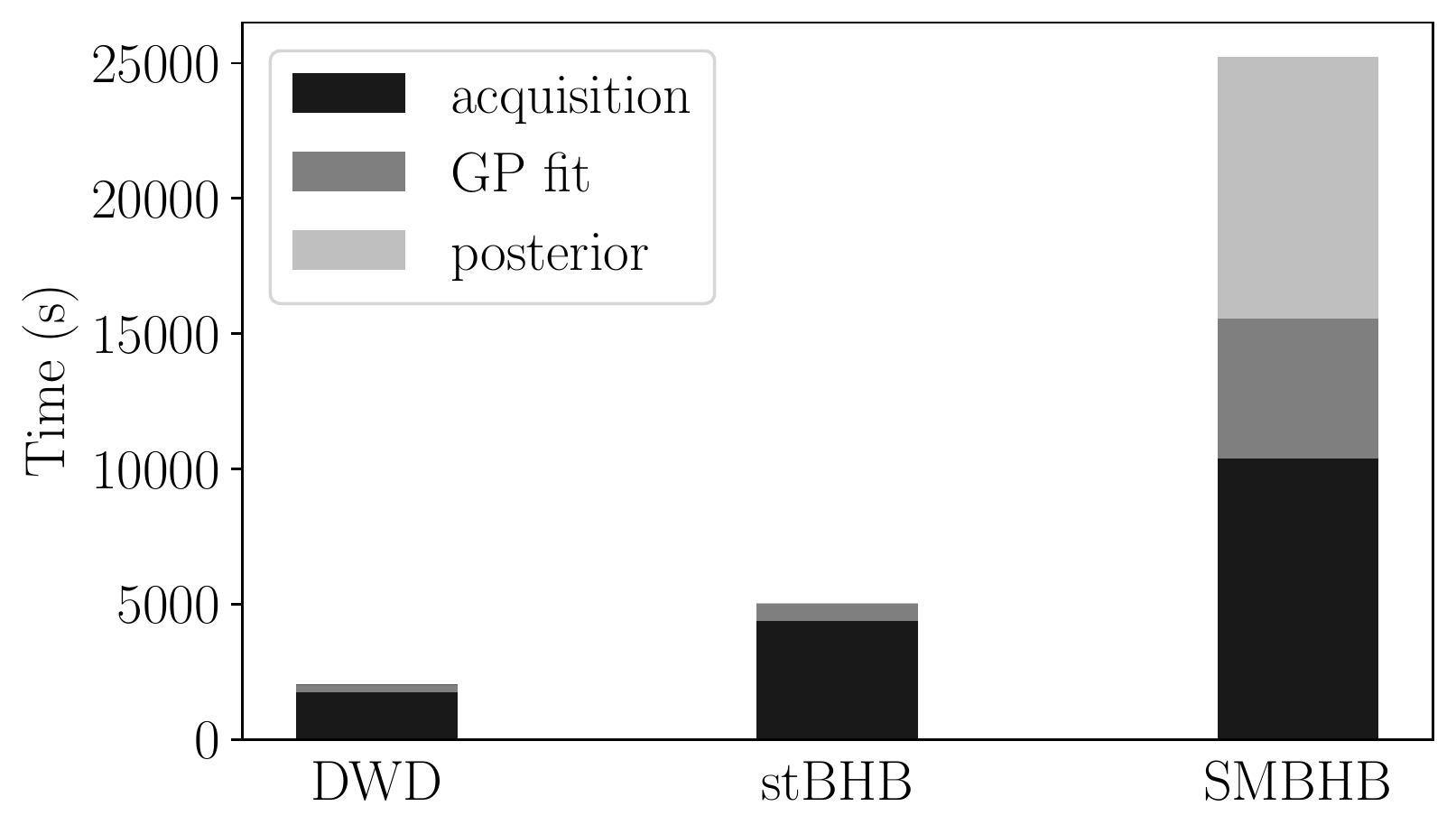}
    \caption{Graph showing the dominant part of the overhead (acquisition and hyperparameter fits of the GPR) vs the total time spent on evaluating the log-posterior. Sub-dominant or non-necessary contributions to \gpry's overhead have been omitted such as determining convergence, checkpointing and the generation of the final MC sample, which typically add up to a few seconds. For the relatively fast to evaluate posteriors of the \dwds and \sbhbs the runtime is dominated by \gpry's overhead which -- due to the much lower number of posterior evaluations -- still leads to a speedup over \nessai in the case of the \sbhbs and \smbhbs. For the slow \smbhb posterior, despite only roughly $1/3$ of the time being spent on posterior evaluations, the large reduction in their number with respect to \nessai still leads to a significant speedup (see \cref{fig:sampling_times}).}
    \label{fig:gpry_overhead_vs_post}
\end{figure}

\cref{tab:gpry_settings} shows the values of the \gpry settings adapted to this study, as motivated in \cref{sec:inf}. An in-depth explanation of the meaning of each setting can be found in \gpry's documentation\footnote{\url{https://gpry.readthedocs.io}}. In \cref{fig:gpry_overhead_vs_post} we show a breakdown of the computation costs of the \gpry runs into posterior evaluation time and overhead from the two computationally expensive steps of the Bayesian optimization loop.

\begin{table*}[ht]
  \begin{tabular}{|l|l|c|c|c|}
    \hline
    \textbf{Setting} & \textbf{Description} & \textbf{\dwd} & \textbf{\sbhb} & \textbf{\smbhb} \\ \hline\hline
    \texttt{noise\_level} & Expected level of numerical noise & $0.1$ & $4$ & $2$ \\\hline
    \texttt{inf\_threshold} & Cutoff in log-posterior of the SVM classifier & $20\sigma$ & $10\sigma$ & $30\sigma$ \\ \hline
    \texttt{fit\_full\_every} & Number of iterations between GPR hyperparameter optimizations & $2$ & $2$ & $1$ \\ \hline
    \texttt{n\_restarts\_optimizer} & Number of restarts per GPR hyperparameter optimization & $2d$ & $d$ & $4d$ \\\hline
    \texttt{mc\_every} & Number of iterations between NS runs to generate proposals & $5$ & $3$ & $3$ \\ \hline
    \texttt{n\_points\_per\_acq} & Number of Kriging steps determining the proposal batch size & $7$ & $9$ & $8$ \\ \hline
    \texttt{trust\_region\_nstd} & Cutoff in log-posterior for defining the trust region & --- & $3\sigma$ & $3\sigma$ \\ \hline
    \texttt{trust\_region\_factor} & Enlargement factor of the trust region & --- & $2.5$ & $2.5$ \\ \hline
  \end{tabular}
  \caption{\label{tab:gpry_settings}
    Non-default settings for \gpry, as discussed in \cref{sec:inf}. In the parameter values, a number followed by $d$ is multiplied by the dimensionality of the problem, whereas one followed by $\sigma$ represents the difference between the log-posterior of the cutoff and that of the best training sample as the equivalent number of 1-dimensional standard deviations, i.e.\ $2\sigma$ represents the log-posterior difference from the top of the distribution of a $d$-dimensional Gaussian that leaves $95\%$ of the mass above it.
  }
\end{table*}

\section{Corner plots}

In this appendix, we show some corner plots for the sources studied in \cref{sec:results}. In the upper part of each plot we furthermore show the sampling locations of \gpry, omitting the samples that are far away from the mode (typically a few percent). The contours for \gpry are obtained by sampling the surrogate model with an MCMC.

\begin{figure}[ht]
    \centering
    \includegraphics[width=\columnwidth]{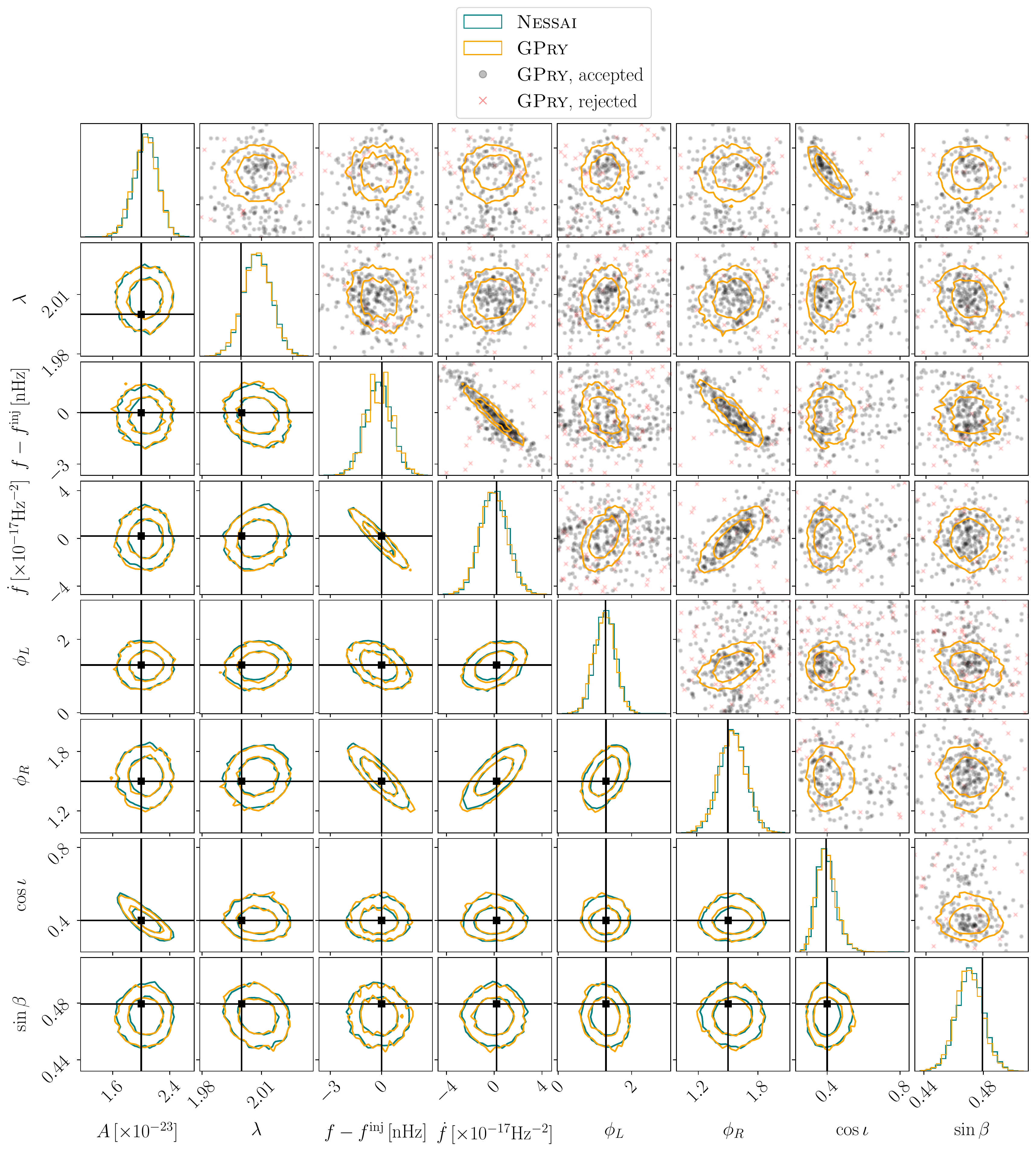}
    \caption{Corner plot comparing \nessai to \gpry inference on the \dwd source with the parameters specified in \cref{tab:WD_params} for the run with the median JS divergence $\djs=0.0043$ (see \cref{fig:WD_js} for the distribution). The number of likelihood evaluations for \gpry was $\approx 500$ (shown in the upper triangle, missing a few percent that would fall outside the ranges of the plot), and for \nessai it was $\approx136500$. The 2d contour levels show the $68\%$ and $95\%$ CL constraints. On the upper triangular we show the locations where \gpry has evaluated the true posterior distribution. The gray dots represent accepted samples (samples that are used to train the GPR), while the red crosses are rejected (used to train the SVM classifier).
    }
    \label{fig:WD_corner_noisy}
\end{figure}

\begin{figure}[ht]
    \centering
    \includegraphics[width=\columnwidth]{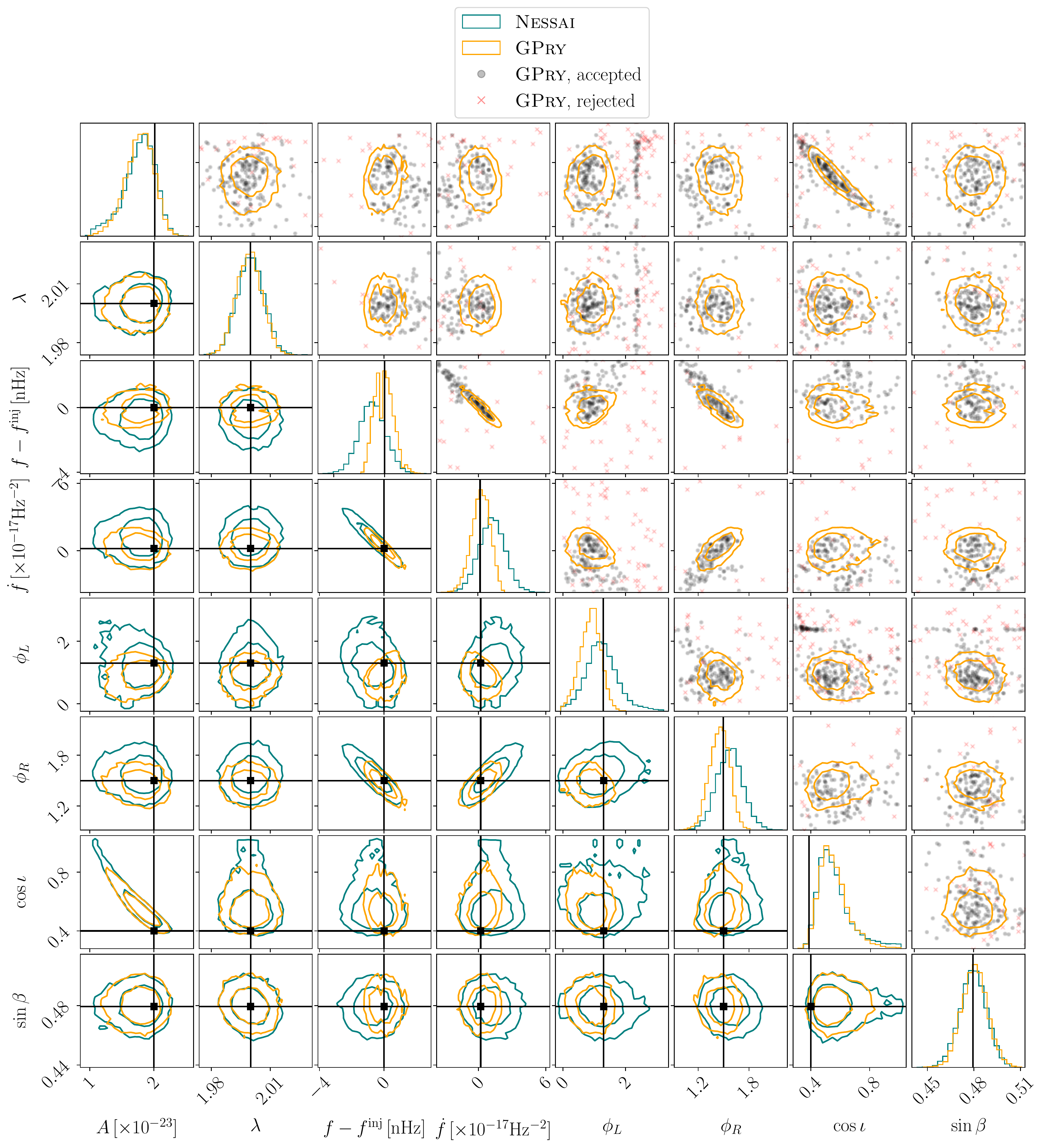}
    \caption{
      Same as \cref{fig:WD_corner_noisy}, comparing \nessai to \gpry inference on the \dwd source with the parameters specified in \cref{tab:WD_params} for the run with the highest JS divergence $\djs=0.12$ (see \cref{fig:WD_js} for the distribution). The number of likelihood evaluations for \gpry was $\approx350$ (shown in the upper triangle), and for \nessai it was $\approx146000$.
    }
    \label{fig:WD_corner_noisy_failed}
\end{figure}

\begin{figure}
    \centering
    \includegraphics[width=\columnwidth]{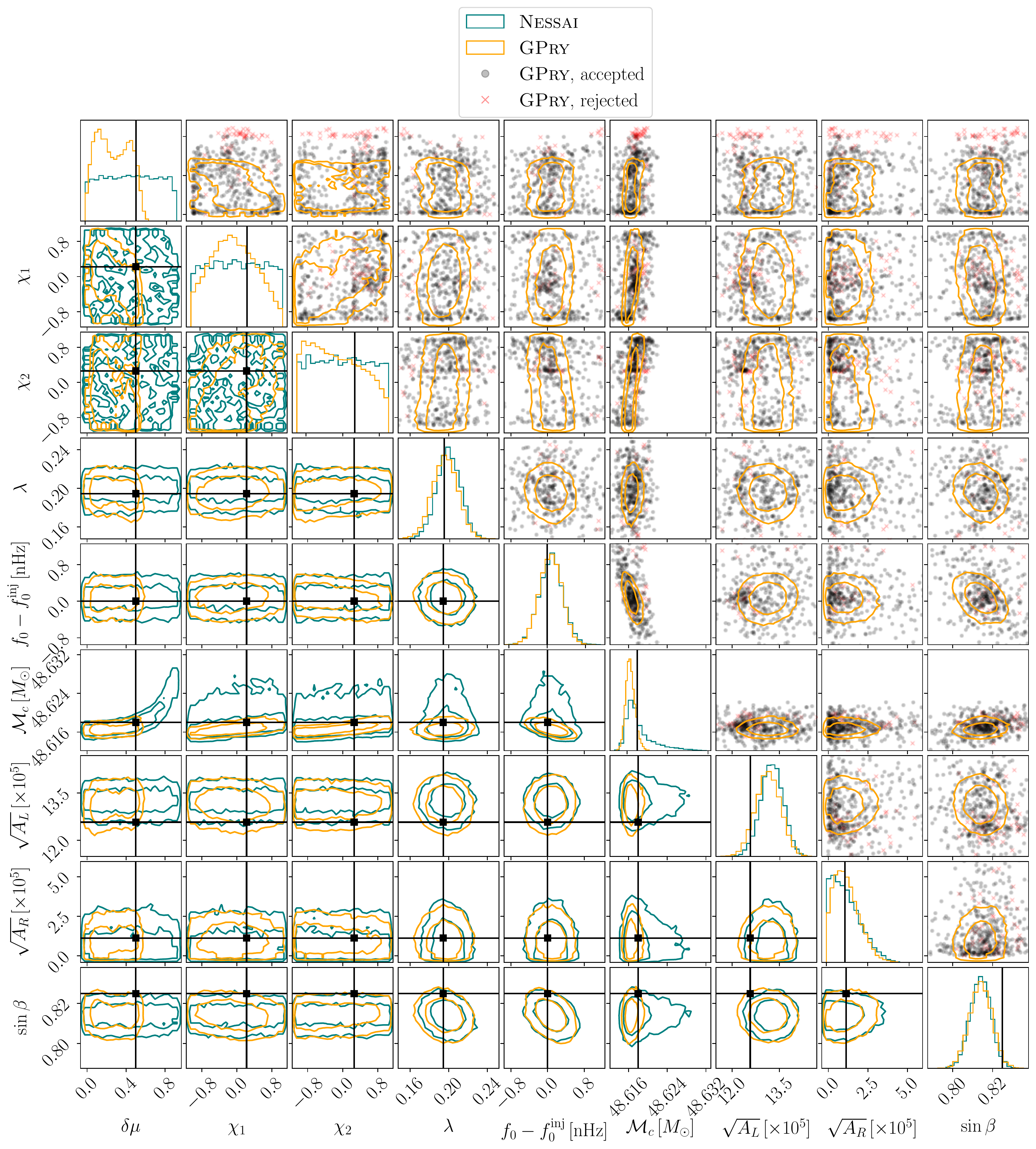}
    \caption{Same as \cref{fig:WD_corner_noisy}, comparing \nessai to \gpry inference on the \sbhb source with the parameters specified in \cref{tab:SOBBH_params} for the run with the median JS divergence $\djs=0.19$ (see \cref{fig:SOBBH_js} for the distribution). The number of likelihood evaluations for \gpry was $\approx450$, and for \nessai it was $\approx 23484500$.
    }
    \label{fig:SOBBH_corner_noisy}
\end{figure}

\begin{figure}
    \centering
    \includegraphics[width=\columnwidth]{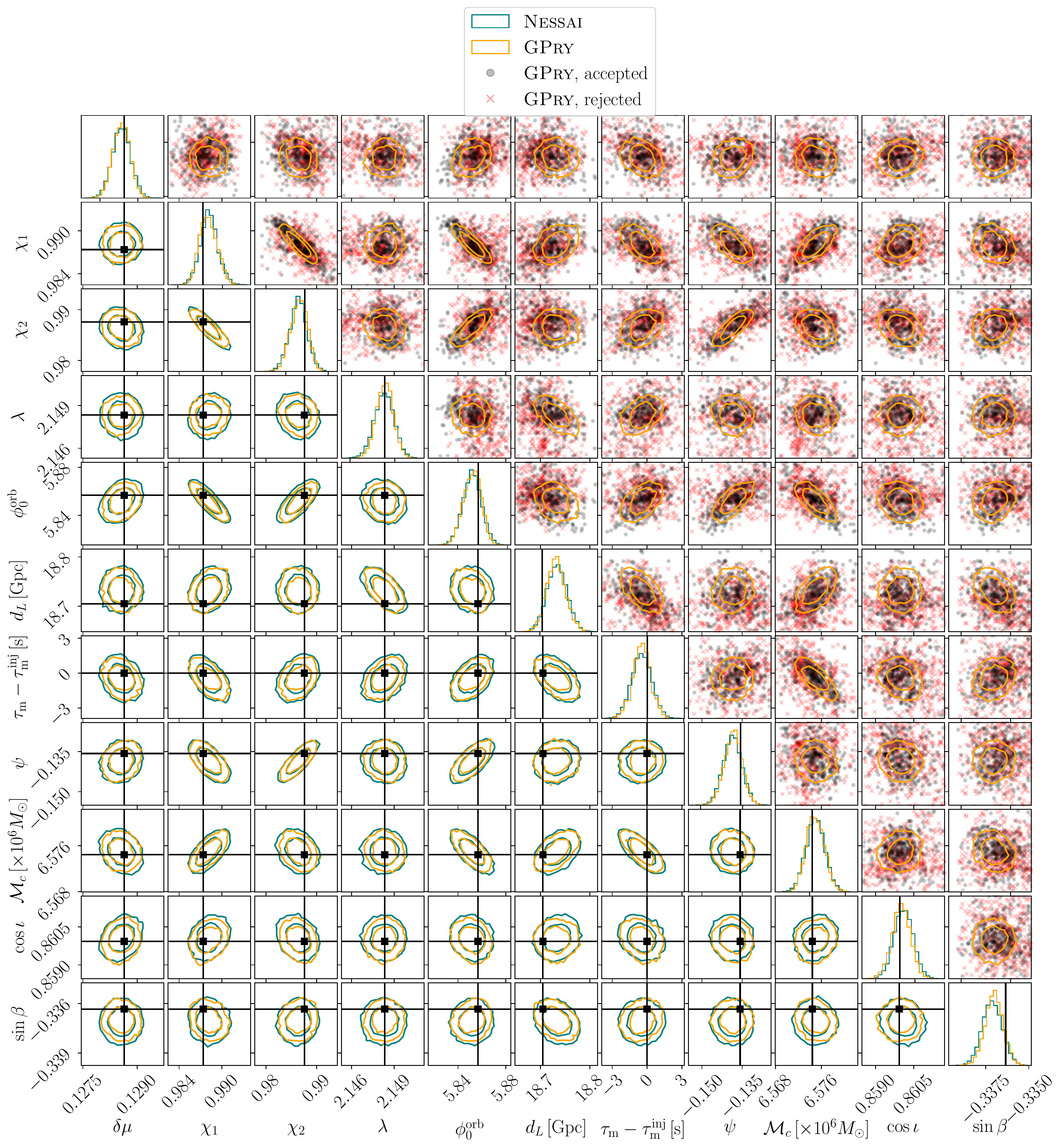}
    \caption{Same as \cref{fig:WD_corner_noisy}, comparing \nessai to \gpry inference on the \smbhb source with the parameters specified in \cref{tab:SMBBH_params} for the run with the median JS divergence $\djs=0.048$ (see \cref{fig:SMBBH_js} for the distribution). The number of likelihood evaluations for \gpry was $\approx850$, and for \nessai it was $\approx 207000$ with 500 live points (used for the $\djs$ calculation and the PP plot), and $\approx567000$ for the high resolution run (2000 live points) whose contours are shown.}
    \label{fig:SMBBH_corner_noisy}
\end{figure}

\begin{figure}
    \centering
    \includegraphics[width=\columnwidth]{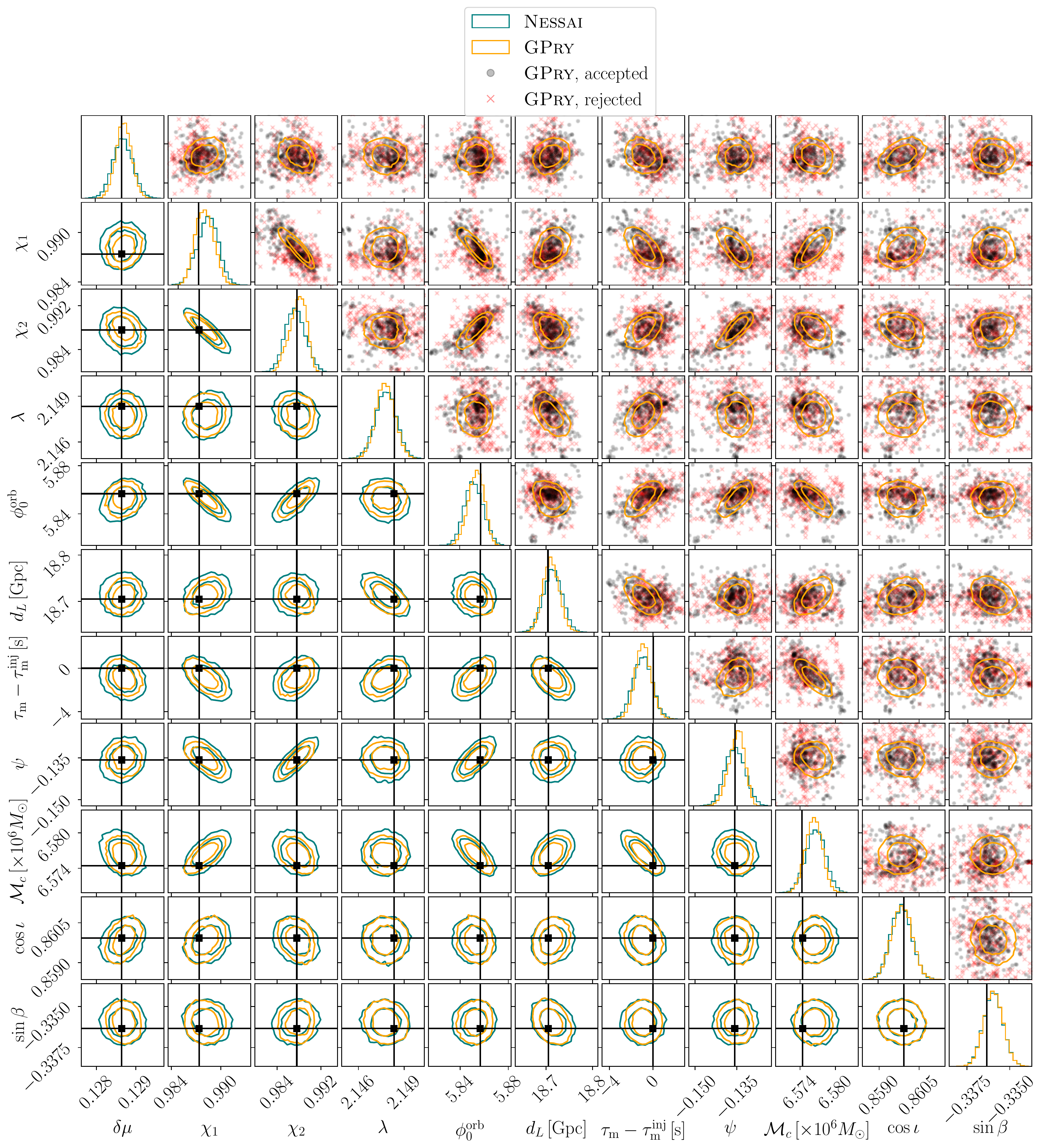}
    \caption{Same as \cref{fig:WD_corner_noisy}, comparing \nessai to \gpry inference on the \smbhb source with the parameters specified in \cref{tab:SMBBH_params} for the run with the highest JS divergence $\djs=0.078$ (see \cref{fig:SMBBH_js} for the distribution). The number of likelihood evaluations for \gpry was $\approx650$, and for \nessai it was $\approx 198500$ at low resolution (500 live points), $\approx 364000$ at high resolution (2000 live points), whose contours are shown.
    }
    \label{fig:SMBBH_corner_noisy_notsogood}
\end{figure}
\end{document}